\documentclass[a4paper,11pt]{article}
\usepackage[utf8]{inputenc}
\usepackage{hyperref}
\usepackage{amsfonts}
\usepackage{amsmath}
\usepackage{amssymb}
\usepackage{indentfirst}
\usepackage{graphicx}
\usepackage{cite}
\usepackage{color}
\usepackage{xcolor}
\usepackage[symbol]{footmisc}

\usepackage{breqn}

\usepackage{setspace}%

\hypersetup{
colorlinks=true,
citecolor=black,
linkcolor=black,
filecolor=black,
urlcolor=black,
}
\numberwithin{equation}{section}
\allowdisplaybreaks
\makeatletter

\begin{document}

\begin{titlepage}

\rightline{  }

\baselineskip=24pt

\begin{center}
\textbf{\LARGE{Extended Bargmann FDA and non-relativistic gravity}}
\par\end{center}{\LARGE \par}

\begin{center}
	\vspace{0.5cm}
	\textbf{Ariana Muñoz,}$^{a,b,}$\footnote[1]{ariana.munoz@uautonoma.cl} \textbf{Gustavo Rubio,}$^{b,}$\footnote[2]{gustavo.rubio@uautonoma.cl} \textbf{Sebastián Salgado}$^{c,}$\footnote[3]{ssalgador@gestion.uta.cl} 
	\small
	\\[4mm]
	$^{a}$\textit{Departamento de \'Optica, Facultad de F\'isica, Universidad Complutense, 28040 Madrid, Spain}
	\\[2mm]
	$^{b}$\textit{Facultad de Ingenier\'ia, Universidad Aut\'onoma de Chile, 5 Poniente 1670, Talca, Chile}
	\\[2mm]
	$^{c}$\textit{Instituto de Alta Investigaci\'on, Universidad de Tarapac\'a, Casilla 7D,
Arica, Chile}

\vskip 7pt
	\footnotesize
	\par\end{center}
\vskip 20pt
%\begin{abstract}
\begin{abstract}
\noindent In this paper we consider the construction of a free differential algebra as
an extension of the extended Bargmann algebra in arbitrary dimensions. This is
achieved by introducing a new Maurer-Cartan equation for a three-form gauge
multiplet in the adjoint representation of the extended Bargmann algebra. The
new Maurer-Cartan equation is provided of non-triviality by means of the
introduction of a four-form cocycle, representative of a Chevalley-Eilenberg
cohomology class. We derive the corresponding dual $L_{\infty}$ algebra and,
by using the formalism of non-linear realizations, propose a five-dimensional
gauge invariant action principle. Then, we derive the corresponding equations
of motion and study how the presence of the three-form gauge fields and the
four-cocycle modify the corresponding non-relativistic dynamics.

	\end{abstract}
\end{titlepage}\newpage {} %{\baselineskip=12pt{} }

\noindent\rule{162mm}{0.4pt}
\tableofcontents
%\noindent\makebox[\linewidth]{\rule{\paperwidth}{0.4pt}}
\noindent\rule{162mm}{0.4pt}

\section{Introduction}

In recent years, several models have been proposed to describe the dynamics of
non-relativistic gravity, especially in the three-dimensional case. First
attempts of obtaining non-relativistic algebras were made in refs.
\cite{Cartan:1923zea,Cartan:1924yea}, where a non-relativistic version of the
Poincar\'{e} algebra was derived by splitting its Lorentz index into space and
time components, and then performing an In\"{o}n\"{u}-Wigner contraction that,
from a physical point of view, corresponds to the limit in which the speed of
light becomes infinite. Further developments and applications of the Galilean
algebra and other non-relativistic symmetries
\cite{Bacry:1968zf,Bacry:1970ye,Trautman,Ehlers} have gained a growing
attention in recent years due to their relevance in several fields of
theoretical physics, such as condensed matter, holography
\cite{Son:2008ye,Bagchi:2009my,Hartong:2015zia,Bagchi:2009pe,Taylor:2015glc}
and the study of the quantum Hall effect
\cite{Son:2013rqa,Hoyos:2011ez,Geracie:2015dea,Gromov:2015fda}. Once
introduced a non-relativistic symmetry, the natural step forward would be the
construction of a gauge invariant action principle from where the dynamics of
non-relativistic gravity would be derived. However, such action principle
cannot be constructed from the Galilei algebra, since the requirement of the
Poisson equation as resulting dynamics is not compatible with the field
content of its gauging and requires a different structure
\cite{Hansen:2019pkl,Hansen:2020wqw}. Furthermore, the Galilei algebra
presents a degenerate invariant bilinear tensor, which makes it impossible to
write down a Chern-Simons action in analogy to the well-known formulation of
three-dimensional Einstein gravity as a Chern-Simons theory of the AdS group
\cite{Achucarro:1986uwr,Witten:1988hc}. Further developments shown it is
possible to extend the Galilei algebra though a central extension, which gives
rise to the so-called Bargmann algebra. Although this extension inherits the
mentioned problem of a degenerate invariant tensor, it makes possible to
reformulate non-relativistic gravity in a geometric way by gauging the algebra
and imposing a gauge-covariant ansatz into the Poisson equation
\cite{Andringa:2010it,Andringa:2016sot}. Later studies have shown that a
three-dimensional Chern-Simons action principle can indeed be constructed by
considering an extension of the Bargmann algebra \cite{Papageorgiou:2009zc}.
This algebra, referred as extended Bargmann algebra (EB), presents a
non-degenerate bilinear invariant tensor, which makes possible the writing of
a well-defined Chern-Simons action principle. This gravity theory can be
considered the non-relativistic counterpart of the three-dimensional Einstein
gravity theory in flat spacetime . Similar results were carried out in the
context of non-relativistic supergravity
\cite{Bergshoeff:2016lwr,Ozdemir:2019tby}. This makes EB gravity particularly
interesting due to its relevance in the study of non-relativistic holography
\cite{Afshar:2015aku}. Although the three-dimensional EB Chern-Simons theory
is a well-defined gauge theory for gravity, it does not fulfill the
requirement of leading to the Poisson equation.\ This feature has been the
motivation in the search of enlarged symmetries, known as Newtonian algebras
\cite{Ozdemir:2019orp}. These are extensions of the EB algebra that also
present non-degenerate invariant tensors, and consequently well-defined
Chern-Simons actions whose equations of motion lead to the Poisson equation.
These studies have been carried out in the context of three-dimensional
gravity, while the study of similar symmetries and gravity theories in higher
dimensions remains mostly unexplored. Of particular interest is the case of
five-dimensional theories, where Chern-Simons forms can also be defined and
therefore, the existence of pure topological theories is allowed. However,
their construction is in general more challenging due to the problem of
obtaining rank-three invariant tensors for the five-dimensional versions of
the gauge algebras. Moreover, the extended Newtonian non-relativistic algebras
usually present a more complicated structure \cite{Concha:2022jdc}. It becomes
therefore relevant to explore new formalisms for the construction of abstract
gauge invariant theories in the context of non-relativistic gravity. For this
purpose, it is noteworthy that Lie algebras can be generalized by means of
their own topological properties. This is the case of free-differential
algebras (FDAs), which are defined as extensions of the dual formulation of
Lie algebras in terms of Maurer-Cartan differential equations for
left-invariant one-forms. Such extensions are performed by considering a basis
of differential forms not only composed by one-forms but by differential forms
of different degrees. This feature enlarges the field content of the
corresponding gauge theories that emerge from them, allowing the construction
of topological theories in more dimensionalities. FDAs have been useful in the
formulation of gauge theories for supergravity in higher-dimensions, allowing
to formulate them in a geometric way
\cite{DAuria:1980cmy,DAuria:1982uck,castellani1991supergravity}. Moreover, it
has been proved that they also allow the existence of Chern-Simons forms in
which that couple the one-forms of standard Lie gauge theories with
independent higher-degree forms
\cite{Banados:1997qs,Antoniadis:2013jja,Konitopoulos:2014owa,Salgado:2017goq,Salgado:2021eli}%
. The main obstacle for the construction of these theories is the finding new
FDAs that non-trivially extend particular Lie algebras, and the later writing
of their corresponding invariant tensors. However, once obtained a FDA of
interest, it is possible to overcome the second difficulty by constructing
non-topological gauge theories, such as the mentioned geometric Lagrangians
or, as it is considered in this work, the formalism of non-linear realizations.

In this article we address two related goals. The first one consisting in
finding a new FDA that non-trivially extends the EB Lie algebra, thus
extending the symmetry and enhancing the field content of one-forms by the
inclusion of three-forms. The second objective is to formulate standard
Newtonian gravity as a gauge invariant theory by means of the formalism of
non-linear realizations. In this framework, we consider two cases: the EB Lie
algebra and its corresponding FDA extension. The study of the second case will
allow us to study how the inclusion of higher-degree forms modifies the
equations of motion of the standard theory. The paper is structured as
follows: In section 2, we present a short review on the theory on FDAs, with
emphasis on the particular case known as FDA1. In section 3, we introduce a
FDA extension of the EB algebra and present its dual formulation. In section
4, we propose a non-linear realization of the EB algebra and its FDA
extension. We make use of them to formulate standard 4D Newtonian gravity as
an off-shell gauge invariant theory of the EB algebra, and a five-dimensional
gauge invariant action that couples non-relativistic gravity with three-forms.

\section{Free differential algebras}

Lie algebras are formulated as vector spaces endowed with antisymmetric
bilinear products satisfying the Jacobi identity. These algebras can be
equivalently formulated by considering a set of left-invariant one-forms in
the group manifold satisfying the so-called Maurer-Cartan equations, which are
equivalent to the statement of zero gauge curvature in the corresponding group
manifold. FDAs are generalizations of the dual formulation of Lie algebras,
which appear by considering a basis of differential forms of different degrees
defined in a smooth manifold, thus allowing the formulation of classical gauge
theories with higher-degree differential forms as gauge fields
\cite{DAuria:1980cmy,castellani1991supergravity}. The simplest case, known as
FDA1, consists of a differential algebra carrying a one-form $A^{A}$, such as
it happens with Lie algebras, and a $p$-form $B^{I}$
\cite{Castellani:1995gz,Castellani:2005vt,Castellani:2013mka}. The
corresponding Maurer-Cartan equations are defined as\footnote{Note that, if
$p=2$, a term of the form $C_{i}^{A}B^{i}$ would be allowed in eq.
(\ref{mc1}). This term is not included in the simplest case known as FDA1.
FDAs that share this feature are called minimal.}:%
\begin{align}
R\left(  A^{A}\right)   &  \equiv\mathrm{d}A^{A}+\frac{1}{2}C_{BC}^{A}%
A^{B}A^{C}=0,\label{mc1}\\
H\left(  B^{I}\right)   &  \equiv\mathrm{d}B^{I}+C_{AJ}^{I}A^{A}B^{J}+\frac
{1}{\left(  p+1\right)  !}C_{A_{1}\cdots A_{p+1}}^{I}A^{A_{1}}\cdots
A^{A_{p+1}}=0. \label{mc2}%
\end{align}
At this point, the structure constants $C_{BC}^{A}$, $C_{AJ}^{I}$ and
$C_{A_{1}\cdots A_{p+1}}^{I}$ are introduced initially without imposing any
conditions. However, in order to have a well-defined differential algebra in
the soft manifold on which these differential forms are introduced, the
exterior derivative operator must satisfy $\mathrm{d}^{2}=0$. This requirement
imposes conditions on the structure constants that are summarized as follows%
\begin{align}
C_{B[C}^{A}C_{BC]}^{A}  &  =0,\label{jac1}\\
C_{AJ}^{I}C_{BK}^{J}-C_{BJ}^{I}C_{AK}^{J}  &  =C_{AB}^{C}C_{CK}^{I}%
,\label{jac2}\\
2C_{[A_{1}|J}^{I}C_{|A_{2}\cdots A_{p+2}]}^{J}-\left(  p+1\right)
C_{B[A_{1}\cdots A_{p}}^{I}C_{A_{p+1}A_{p+2]}}^{B}  &  =0. \label{jac3}%
\end{align}
Eqs. (\ref{jac1})-(\ref{jac3}) are the generalization of the Jacobi identity
for FDA1. Notice that eq. (\ref{jac1}) is the standard Jacobi identity of a
Lie algebra, and the structure constants $C_{BC}^{A}$ are antisymmetric in the
lower indices by virtue of eq. (\ref{mc1}). This shows that FDA1 presents a
Lie subalgebra. Moreover, eq. (\ref{jac2}) shows that the structure constants
$C_{AJ}^{I}$ form a representation of that Lie subalgebra, which means that,
when extending a Lie algebra to a FDA1 the index $I$ of the $p$-form must be a
representation index
\cite{Castellani:1995gz,Castellani:2005vt,Castellani:2013mka,Castellani:2006jg}%
. The fact that the structure constants $C_{AJ}^{I}$ form a representation
implies that the new Marurer-Cartan equations for the higher-degree forms can
be written in terms of the covariant derivative of the Lie algebra, as follows%
\begin{equation}
\nabla B^{I}+\Omega^{I}=0, \label{mc2b}%
\end{equation}
with%
\begin{equation}
\Omega^{I}=\frac{1}{\left(  p+1\right)  !}C_{A_{1}\cdots A_{p+1}}^{I}A^{A_{1}%
}\cdots A^{A_{p+1}},
\end{equation}
where the covariant derivative is defined in terms of the chosen
representation as $\nabla B^{I}=\mathrm{d}B^{J}+C_{AJ}^{I}A^{A}B^{J}$.
Finally, the third structure constant $C_{A_{1}\cdots A_{p+1}}^{I}$ is always
antisymmetric in the lower indices. Eq. (\ref{jac3}) turns out to be
equivalent to the statement $\nabla\Omega^{I}=0$. This shows that the $\left(
p+1\right)  $-form $\Omega^{I}$ introduced in eq. (\ref{jac3}) must be a
cocycle of the Lie subalgebra. Cocycles are classified in Chevalley-Eilenberg
cohomology classes of differential forms that are covariantly closed and
non-covariantly exact \cite{Chevalley:1948zz}. Thus, two cocycles $\Omega$ and
$\Omega^{\prime}$ belong to the same cohomology if they differ in a
covariantly exact form (as $\Omega^{\prime}=\Omega+\nabla\varphi$ for some
$p$-form $\varphi$ defined in terms of $A^{A}$ and $B^{I}$). Similarly, two
algebras of the FDA1 type are equivalent if they are constructed with cocycles
belonging to the same cohomology class. Moreover, if a FDA1 is constructed
with a trivial cocycle (i.e., a covariantly exact cocycle), such algebra is
equivalent to an algebra constructed without any cocycle, since its
contribution to the Maurer-Cartan equations can be reabsorbed by a
redefinition of the potential $B^{I}$. From an inspection of eqs. (\ref{mc1})
and (\ref{mc2}), it follows that a general Lie algebra can be extended to FDA1
by introducing a $p$-form in a given representation, if and only if a $\left(
p+1\right)  $-cocycle in the same representation exists. The use of FDA1 in
the construction of gauge theories for gravity allows to extend the gauge
principle by introducing $p$-form gauge fields.

\section{Extended Bargmann FDA}

The EB Lie algebra can be obtained as an expansion of the Poincar\'{e} algebra
\cite{deAzcarraga:2019mdn,Concha:2023bly}. In fact, let us consider the
$D$-dimensional Poincar\'{e} algebra $\mathfrak{iso}\left(  D-1,1\right)  $,
whose generators $\left\{  \mathbf{\hat{P}}_{a},\mathbf{\hat{J}}_{ab}\right\}
$ satisfy the following commutation relations%
\begin{align}
\left[  \mathbf{\hat{J}}_{ab},\mathbf{\hat{J}}_{cd}\right]   &  =\eta
_{bc}\mathbf{\hat{J}}_{ad}+\eta_{ad}\mathbf{\hat{J}}_{bc}-\eta_{ac}%
\mathbf{\hat{J}}_{bd}-\eta_{bd}\mathbf{\hat{J}}_{ac},\\
\left[  \mathbf{\hat{J}}_{ab},\mathbf{\hat{P}}_{c}\right]   &  =\eta
_{bc}\mathbf{\hat{P}}_{a}-\eta_{ac}\mathbf{\hat{P}}_{b}.
\end{align}
Here $a,b=0,...,D-1$ are Lorentz indices, raised and lowered with the
$D$-dimensional Minkowski metric $\eta_{ab}=\mathrm{diag}\left(
-,+,\overset{D-1}{\cdots},+\right)  $. Let us also consider the semigroup
$S_{E}^{\left(  2\right)  }=\left\{  \lambda_{0},\lambda_{1},\lambda
_{2},\lambda_{3}\right\}  $, which is endowed with the multiplication rule%
\begin{equation}
\lambda_{\alpha}\lambda_{\beta}=\left\{
\begin{tabular}
[c]{ll}%
$\lambda_{\alpha+\beta}\,\,$ & $\mathrm{if}\,\,\,\,\alpha+\beta\leq3\,,$\\
$\lambda_{3}$ & $\mathrm{otherwise}.$%
\end{tabular}
\ \ \right.
\end{equation}
Note that $\lambda_{3}$ is a zero element. We decompose this semigroup as
$S_{E}^{\left(  2\right)  }=S_{0}\cup S_{1}$ with $S_{0}=\left\{  \lambda
_{0},\lambda_{2},\lambda_{3}\right\}  $ and $S_{1}=\left\{  \lambda
_{1},\lambda_{3}\right\}  $. Moreover, by splitting the Lorentz index into
space and time components as $a=\left(  0,i\right)  $, the Poincar\'{e}
algebra can be decomposed as $\mathfrak{iso}\left(  D-1,1\right)  =V_{0}\oplus
V_{1}$ with $V_{0}=\mathrm{span}\left\{  \mathbf{\hat{J}}_{ij},\mathbf{\hat
{P}}_{0}\,\right\}  $ and $V_{1}=\mathrm{span}\left\{  \mathbf{\hat{J}}%
_{0i},\mathbf{\hat{P}}_{i}\,\right\}  $. According to the definitions of ref.
\cite{Izaurieta:2006zz}, both decompositions are in resonance, which allows to
introduce the resonant $S_{E}^{\left(  2\right)  }$-expanded algebra, defined
by $\mathfrak{G}_{R}=\left(  S_{0}\times V_{0}\right)  \oplus\left(
S_{1}\times V_{1}\right)  $. Moreover, the presence of the zero element in the
semigroup allows us to extract a reduced algebra from $\mathfrak{G}_{R}$ by
considering $\lambda_{3}\times\left\{  \mathbf{\hat{P}}_{a},\mathbf{\hat{J}%
}_{ab}\right\}  =0$. Thus, by defining the expanded generators%
\begin{align}
\mathbf{J}_{ij}  &  =\lambda_{0}\mathbf{\hat{J}}_{ij},\qquad\mathbf{H}%
=\lambda_{0}\mathbf{\hat{P}}_{0}\,,\qquad\mathbf{G}_{i}=\lambda_{1}%
\mathbf{\hat{J}}_{0i},\nonumber\\
\mathbf{M}  &  =\lambda_{2}\mathbf{\hat{P}}_{0}\,,\qquad\mathbf{P}_{i}%
=\lambda_{1}\mathbf{\hat{P}}_{i}\,,\qquad\mathbf{S}_{ij}=\lambda
_{2}\mathbf{\hat{J}}_{ij}\,, \label{expansion}%
\end{align}
one finds that they span the EB algebra, satisfying the following commutators:%
\begin{align}
\left[  \mathbf{J}_{ij},\mathbf{J}_{kl}\right]   &  =\delta_{ik}%
\mathbf{J}_{lj}-\delta_{il}\mathbf{J}_{kj}+\delta_{jl}\mathbf{J}_{ki}%
-\delta_{jk}\mathbf{J}_{li},\nonumber\\
\lbrack\mathbf{J}_{ij},\mathbf{G}_{k}]  &  =\delta_{kj}\mathbf{G}_{i}%
-\delta_{ki}\mathbf{G}_{j},\nonumber\\
\text{ \ \ \ }[\mathbf{J}_{ij},\mathbf{P}_{k}]  &  =\delta_{kj}\mathbf{P}%
_{i}-\delta_{ki}\mathbf{P}_{j},\nonumber\\
\lbrack\mathbf{G}_{i},\mathbf{P}_{j}]  &  =\delta_{ij}\mathbf{M},\nonumber\\
\text{ \ \ \ }[\mathbf{G}_{i},\mathbf{H}]  &  =\mathbf{P}_{i},\nonumber\\
\lbrack\mathbf{G}_{i},\mathbf{G}_{j}]  &  =\mathbf{S}_{ij},\text{\ \ \ }%
\nonumber\\
\left[  \mathbf{J}_{ij},\mathbf{S}_{kl}\right]   &  =\delta_{ik}%
\mathbf{S}_{lj}-\delta_{il}\mathbf{S}_{kj}+\delta_{jl}\mathbf{S}_{ki}%
-\delta_{jk}\mathbf{S}_{li}. \label{eb17}%
\end{align}
Thus, the EB algebra appears as a $S_{E}^{\left(  2\right)  }$-expansion of
the Poincar\'{e} algebra without the need of an Inonu-Wigner contraction
\cite{deAzcarraga:2019mdn,Concha:2023bly}. From eqs. (\ref{eb17}), we see that
the generators $\mathbf{S}_{ij}$ extend the Bargmann algebra, which is spanned
by the set $\left\{  \mathbf{J}_{ij}\mathbf{,G}_{i}\mathbf{,P}_{i}%
\mathbf{,H,M}\right\}  $. Similarly, the Bargmann algebra appears as a central
extension of the Galilei algebra spanned by $\left\{  \mathbf{J}%
_{ij}\mathbf{,G}_{i}\mathbf{,P}_{i}\mathbf{,H}\right\}  $.

\subsection{Cohomology class}

Let us now address the goal of obtaining an extension of the EB algebra. For
this purpose, let us introduce the one-form gauge connection evaluated in
$\mathfrak{iso}\left(  D-1,1\right)  $%
\begin{equation}
\mathbf{\hat{A}}=\hat{e}^{a}\mathbf{\hat{P}}_{a}+\frac{1}{2}\hat{\omega}%
^{ab}\mathbf{\hat{J}}_{ab}.
\end{equation}
The Poincar\'{e} algebra presents the following cocycle
\begin{equation}
\mathbf{\hat{\Omega}}=\hat{e}^{a}\hat{\omega}^{bc}\hat{\omega}_{bd}\hat
{\omega}_{\text{ \ }c}^{d}\mathbf{\hat{P}}_{a}. \label{cocycleP}%
\end{equation}
Notice that $\mathbf{\hat{\Omega}}$ is a four-cocycle defined in the adjoint
representation of $\mathfrak{iso}\left(  D-1,1\right)  $, which allows to
write it as a linear combinations of the generators of the algebra
\cite{Salgado:2021eli,Salgado:2023owk}. It is direct to verify that
$\mathbf{\hat{\Omega}}$ satisfies covariant closure, i.e.,
\begin{equation}
\tilde{\nabla}\mathbf{\hat{\Omega}}=\mathrm{d}\mathbf{\hat{\Omega}}+\left[
\mathbf{\hat{A}},\mathbf{\hat{\Omega}}\right]  =0.
\end{equation}
The non-triviality of the cocycle can be proved according to the following
prescription: one proposes the most general ansatz for a four-cochain
$\mathbf{\Omega}_{\text{General}}$ in the adjoint representation of the
Poincar\'{e} algebra, depending on wedge products and index contractions of
the Maurer-Cartan one-forms of the Lie subalgebra $\left(  \hat{e}^{a}%
,\hat{\omega}^{ab}\right)  $ in a way that preserves Lorentz covariance. The
terms of the ansatz carry arbitrary constants that are fixed by the covariant
closure requirement. Thus, after imposing covariant closure, each independent
constant is associated to an independent cocycle. Then, we propose as a second
ansatz the most general three-cochain $\boldsymbol{\hat{\varphi}}$ that can be
constructed with the same conditions. The covariant derivative of
$\boldsymbol{\hat{\varphi}}$ provides the most general trivial four-cocycle
that can be found in this representation, i.e., $\hat{\Omega}_{\text{Trivial}%
}=\nabla\boldsymbol{\hat{\varphi}}$. By comparing $\mathbf{\hat{\Omega}%
}_{\text{General}}$ with $\mathbf{\hat{\Omega}}_{\text{Trivial}}$, it is
possible to extract the non-trivial components $\mathbf{\hat{\Omega}%
}_{\text{General}}$. This process shows that $\mathbf{\hat{\Omega}}$ is in
fact a non-trivial cocycle, as it cannot be obtained as the covariant
derivative of a three-cochain.

Let us now consider the dual expansion of the Poincar\'{e} algebra
\cite{Izaurieta:2009gc}. This means to perform the same expansion procedure
shown in eqs. (\ref{expansion}) on the left-invariant one-forms of the
Poincar\'{e} algebra, as follows:%
\begin{align}
\hat{e}^{0}  &  =\lambda_{0}\tau+\lambda_{2}m,\text{ \ \ \ \ }\hat{e}%
^{i}=\lambda_{1}e^{i},\nonumber\\
\hat{\omega}^{0k}  &  =\lambda_{1}\omega^{k},\text{ \ \ \ \ }\hat{\omega}%
^{ij}=\lambda_{0}\omega^{ij}+\lambda_{2}s^{ij}.
\end{align}
By applying the dual expansion on eq. (\ref{cocycleP}), we find the following
four-cocycle of the EB algebra%
\begin{equation}
\mathbf{\Omega}=\tau\omega^{jk}\omega_{jl}\omega_{\text{ \ }k}^{l}%
\mathbf{H}+e^{i}\omega^{jk}\omega_{jl}\omega_{\text{ \ }k}^{l}\mathbf{P}%
_{i}+\left(  3\tau\omega_{jl}\omega_{\text{ \ }k}^{l}s^{jk}-3\tau\omega
^{j}\omega^{k}\omega_{kj}+m\omega^{jk}\omega_{jl}\omega_{\text{ \ }k}%
^{l}\right)  \mathbf{M}. \label{ebcocycle}%
\end{equation}
It is easy to verify that $\mathbf{\Omega}$ inherits the covariant closure
property of $\mathbf{\hat{\Omega}}$, regarding the covariant derivative
defined with the extended Bargmann gauge connection, i.e., $\mathrm{d}%
\mathbf{\Omega}+\left[  \mathbf{A},\mathbf{\Omega}\right]  =0$. However,
$\mathbf{\Omega}$ turns out to be composed by two independent cocycles, each
one verifying covariant closure. These are given by%
\begin{align}
\mathbf{\Omega}_{1}  &  =3\left(  \tau\omega_{jl}\omega_{\text{ \ }k}%
^{l}s^{jk}-\tau\omega^{j}\omega^{k}\omega_{kj}\right)  \mathbf{M,}\label{O1}\\
\mathbf{\Omega}_{2}  &  =\tau\omega^{jk}\omega_{jl}\omega_{\text{ \ }k}%
^{l}\mathbf{H+}e^{i}\omega^{jk}\omega_{jl}\omega_{\text{ \ }k}^{l}%
\mathbf{P}_{i}+m\omega^{jk}\omega_{jl}\omega_{\text{ \ }k}^{l}\mathbf{M.}
\label{O2}%
\end{align}
By using the same procedure described above, we find that $\mathbf{\Omega}%
_{1}$ is a trivial cocycle. In fact it is possible to write $\mathbf{\Omega
}_{1}=\nabla\boldsymbol{\varphi}$, with%
\begin{equation}
\boldsymbol{\varphi}=\tau\omega^{ij}s_{ij}\mathbf{M.}%
\end{equation}
In contrast, $\mathbf{\Omega}_{2}$ cannot be written as an exact covariant
derivative, which makes it a non-trivial cocycle, representative of a
Chevalley-Eilenberg cohomology class of the $D$-dimensional EB algebra. It
becomes then possible to introduce a new FDA1 by using the EB Lie algebra as
starting point. To this end, we consider a three-form gauge field in the
adjoint representation of the EB algebra, whose components we denote%
\begin{equation}
B^{A}=\left(  b,c,b^{i},c^{i},b^{ij},c^{ij}\right)  .
\end{equation}
Since the three-form (and the cocycle) is defined in the adjoint
representation of the Lie algebra, its algebraic index takes values in the
same domain as the one-form. Thus, the components $b$, $c$, $b^{i}$, $c^{i}$,
$b^{ij}$ and $c^{ij}$ are associated to the same algebraic sector that $\tau$,
$m$, $e^{i}$, $\omega^{i}$, $\omega^{ij}$ and $s^{ij}$ respectively. We
introduce a new Maurer-Cartan equation for $B^{A}$, which, according to
(\ref{mc2b}) is given by%
\begin{equation}
\nabla B^{A}+\alpha\Omega_{2}^{A}=0.
\end{equation}
Note that we have included a factor $\alpha$. This is a numeric factor that we
introduce for later convenience, and whose value set as $\alpha=0,1$. Thus, we
can later consider the vanishing of the cocycle in the FDA and analyze the
consequences in the resulting gravity theory. The new FDA1, that we denote as
EB-FDA, is therefore defined by two sets of equations:%

\begin{align}
R\left(  \tau\right)   &  \equiv\mathrm{d}\tau=0,\nonumber\\
R\left(  m\right)   &  \equiv\mathrm{d}m+\omega^{i}e_{i}=0,\nonumber\\
R\left(  e^{i}\right)   &  \equiv\mathrm{d}e^{i}+\omega^{ij}e_{j}+\omega
^{i}\tau=0,\nonumber\\
R\left(  \omega^{ij}\right)   &  \equiv\mathrm{d}\omega^{ij}+\omega_{\text{
\ }k}^{i}\omega^{kj}=0,\nonumber\\
R\left(  \omega^{i}\right)   &  \equiv\mathrm{d}\omega^{i}+\omega^{ik}%
\omega_{k}=0,\nonumber\\
R\left(  s^{ij}\right)   &  \equiv\mathrm{d}s^{ij}+\omega_{\text{ \ }k}%
^{i}s^{kj}-\omega_{\text{ \ }k}^{j}s^{ki}+\omega^{i}\omega^{j}=0,
\label{mceb1}%
\end{align}
in addition to%
\begin{align}
H\left(  b\right)   &  \equiv\mathrm{d}b+\alpha\tau\omega^{ij}\omega
_{ik}\omega_{\text{ \ }j}^{k}=0,\nonumber\\
H\left(  c\right)   &  \equiv\mathrm{d}c+\omega^{i}b_{i}+b^{i}e_{i}+\alpha
m\omega^{ij}\omega_{ik}\omega_{\text{ \ }j}^{k}=0,\nonumber\\
H\left(  b^{i}\right)   &  \equiv\mathrm{d}b^{i}+\omega_{\text{ \ }j}^{i}%
b^{j}+b\omega^{i}+c^{i}\tau+b_{\text{ \ }j}^{i}e^{j}+\alpha e^{i}\omega
^{jk}\omega_{jl}\omega_{\text{ \ }k}^{l}=0,\nonumber\\
H\left(  c^{i}\right)   &  \equiv\mathrm{d}c^{i}+\omega_{\text{ \ }j}^{i}%
c^{j}+b_{\text{ \ }j}^{i}\omega^{j}=0,\nonumber\\
H\left(  b^{ij}\right)   &  \equiv\mathrm{d}b^{ij}+\omega_{\text{ \ }k}%
^{i}b^{kj}-\omega_{\text{ \ }k}^{j}b^{ki}=0,\nonumber\\
H\left(  c^{ij}\right)   &  \equiv\mathrm{d}c^{ij}+\omega_{\text{ \ }k}%
^{i}c^{kj}-\omega_{\text{ \ }k}^{j}c^{ki}+s_{\text{ \ }k}^{i}b^{kj}-s_{\text{
\ }k}^{j}b^{ki}+\omega^{i}c^{j}-\omega^{j}c^{i}=0. \label{mceb2}%
\end{align}
Eqs. (\ref{mceb1}) define the EB algebra in dual formulation, while
(\ref{mceb2}) is a set of new Maurer-Cartan equations for the new three-form
gauge potentials. By setting $\alpha=0$, we obtain a trivial extension of the
EB algebra in which the three-forms can be decomposed in terms of the one-forms.

\subsection{Dual algebra}

As it happens with Lie algebras, FDAs can be formulated as vector spaces
endowed with linear products. However, the presence of higher-degree forms in
a FDA makes necessary to introduce multilinear products in addition to the
bilinear one of a Lie algebra. These algebras are known as $L_{\infty}$
algebras are non-associative structures that appear in the study of closed
string theory and the Poisson structures of classical gauge theories
\cite{Jurco:2018sby,Lada:1992wc,Lada:1994mn,Barnich:1997ij}. They are
generalizations of Lie algebras that satisfy a Jacobi-like identity for the
multilinear products, known as $L_{\infty}$ identity
\cite{Hohm:2017pnh,Hohm:2017cey}. Consequently, when writing a $L_{\infty}$
algebra as a FDA, this identity is translated into a generalized Jacobi
identities for the FDA structure constant, which take the form of eqs.
(\ref{jac1})-(\ref{jac3}) for the particular case of a FDA1. The $L_{\infty}$
algebra dual to a general FDA1 is defined as a vector space spanned by the set
of vectors $\left\{  \mathbf{t}_{A}\mathbf{,t}_{i}\right\}  $ dual to the
basis of differential forms $\left[  A^{A},B^{I}\right]  $, endowed with the
following bilinear and multilinear products
\cite{Salgado:2021sjo,Salgado:2023owk}:%
\begin{align}
\left[  \mathbf{t}_{A}\mathbf{,t}_{B}\right]   &  =C_{AB}^{C}\mathbf{t}%
_{C},\label{L1}\\
\left[  \mathbf{t}_{A},\mathbf{t}_{I}\right]   &  =C_{AI}^{J}\mathbf{t}%
_{J},\label{L2}\\
\left[  \mathbf{t}_{A_{1}},...,\mathbf{t}_{A_{p+1}}\right]   &  =p!C_{A_{1}%
\cdots A_{p+1}}^{I}\mathbf{t}_{I},\label{L3}\\
\text{others}  &  =0.
\end{align}
From eq. (\ref{jac1}) and (\ref{L1}), it follows that this $L_{\infty}$
algebra has a Lie subalgebra spanned by $\left\{  \mathbf{t}_{A}\right\}  $.
The product in (\ref{L2}) is symmetric or antisymmetric for $p$ even or odd
respectively. The $\left(  p+1\right)  $-linear product in (\ref{L3}) is
always antisymmetric. If a Lie algebra is extended into a FDA1 by introducing
a $p$-form in the adjoint representation, as it is the case of EB-FDA, the
representation index takes the form $I\rightarrow A$, while the generalized
structure constants $C_{AJ}^{I}$ become equivalent to those of the original
Lie algebra, i.e., $C_{AJ}^{I}\rightarrow C_{AB}^{C}$. To avoid confusion with
the generators of the Lie subalgebra, we rename the vectors along the extended
subspace with a bar $\mathbf{t}_{I}\rightarrow\mathbf{\bar{t}}_{A}$, then the
$L_{\infty}$ algebra associated to FDA1 is spanned by the basis vectors
$\left\{  \mathbf{t}_{A},\mathbf{\bar{t}}_{A}\right\}  $. Then, the product in
eq. (\ref{L1}) remains the same, while those in eqs. (\ref{L2}) and (\ref{L3})
take the form%
\begin{align}
\left[  \mathbf{t}_{A}\mathbf{,\bar{t}}_{B}\right]   &  =C_{AB}^{C}%
\mathbf{\bar{t}}_{C},\label{L2adj}\\
\left[  \mathbf{t}_{A_{1}},...,\mathbf{t}_{A_{p+1}}\right]   &  =p!C_{A_{1}%
\cdots A_{p+1}}^{B}\mathbf{\bar{t}}_{B}. \label{L3adj}%
\end{align}
In the following, we will maintain this notation. All vectors along the
extended sector of the algebra will be denoted with bar.

This dual formulation allows us to write down the extended Bargmann FDA from
eqs. (\ref{mceb1}) and (\ref{mceb2}) in terms of bilinear and multilinear
products. Let us recall that in the case the $\mathbf{t}_{A}$ vectors are
decomposed as $\mathbf{t}_{A}=(\mathbf{H,M,P}_{i}\mathbf{,G}_{i}%
\mathbf{,J}_{ij}\mathbf{,S}_{ij})$. Similarly, we denote the basis vectors
spanning the extended algebraic sector as $\mathbf{\bar{t}}_{A}=(\mathbf{\bar
{H},\bar{M},\bar{P}}_{i}\mathbf{,\bar{G}}_{i}\mathbf{,\bar{J}}_{ij}%
\mathbf{,\bar{S}}_{ij})$. Thus, the relations of the $L_{\infty}$ algebra
(dual to EB-FDA) are given by%
\begin{align}
\left[  \mathbf{J}_{ij}\mathbf{,J}_{kl}\right]   &  =\delta_{\lbrack
i[k}\mathbf{J}_{l]j]}, & \lbrack\mathbf{\bar{J}}_{ij}\mathbf{,P}_{k}]  &
=\delta_{k[j}\mathbf{\bar{P}}_{i]},\nonumber\\
\lbrack\mathbf{J}_{ij}\mathbf{,G}_{k}]  &  =\delta_{k[j}\mathbf{G}_{i]}, &
\lbrack\mathbf{\bar{G}}_{i}\mathbf{,P}_{j}]  &  =\delta_{ij}\mathbf{\bar{M}%
},\nonumber\\
\lbrack\mathbf{J}_{ij}\mathbf{,P}_{k}]  &  =\delta_{k[j}\mathbf{P}_{i]}, &
\lbrack\mathbf{G}_{i}\mathbf{,\bar{P}}_{j}]  &  =\delta_{ij}\mathbf{\bar{M}%
},\nonumber\\
\lbrack\mathbf{G}_{i}\mathbf{,P}_{j}]  &  =\delta_{ij}\mathbf{M}, &
\lbrack\mathbf{\bar{G}}_{i}\mathbf{,H}]  &  =\mathbf{\bar{P}}_{i},\nonumber\\
\lbrack\mathbf{G}_{i}\mathbf{,H}]  &  =\mathbf{P}_{i}, & \lbrack\mathbf{G}%
_{i}\mathbf{,\bar{H}}]  &  =\mathbf{\bar{P}}_{i},\nonumber\\
\lbrack\mathbf{G}_{i}\mathbf{,G}_{j}]  &  =\mathbf{S}_{ij}, & \lbrack
\mathbf{\bar{G}}_{i}\mathbf{,G}_{j}]  &  =\mathbf{\bar{S}}_{ij},\nonumber\\
\left[  \mathbf{J}_{ij}\mathbf{,S}_{kl}\right]   &  =\delta_{\lbrack
i[k}\mathbf{S}_{l]j]}, & \left[  \mathbf{\bar{J}}_{ij}\mathbf{,S}_{kl}\right]
&  =\delta_{\lbrack i[k}\mathbf{\bar{S}}_{l]j]},\nonumber\\
\left[  \mathbf{J}_{ij}\mathbf{,\bar{J}}_{kl}\right]   &  =\delta_{\lbrack
i[k}\mathbf{\bar{J}}_{l]j]}, & \left[  \mathbf{J}_{ij}\mathbf{,\bar{S}}%
_{kl}\right]   &  =\delta_{\lbrack i[k}\mathbf{\bar{S}}_{l]j]},\nonumber\\
\lbrack\mathbf{J}_{ij}\mathbf{,\bar{G}}_{k}]  &  =\delta_{k[j}\mathbf{\bar{G}%
}_{i]}, & \left[  \mathbf{H,J}_{ij}\mathbf{,J}_{kl}\mathbf{,J}_{mn}\right]
&  =(3!)^{2}\alpha\delta_{\lbrack ij,kl,mn]}\mathbf{\bar{H},}\nonumber\\
\lbrack\mathbf{\bar{J}}_{ij}\mathbf{,G}_{k}]  &  =\delta_{k[j}\mathbf{\bar{G}%
}_{i]}, & \left[  \mathbf{P}_{h}\mathbf{,J}_{ij}\mathbf{,J}_{kl}%
\mathbf{,J}_{mn}\right]   &  =(3!)^{2}\alpha\delta_{\lbrack ij,kl,mn]}%
\mathbf{\bar{P}}_{h},\nonumber\\
\lbrack\mathbf{J}_{ij}\mathbf{,\bar{P}}_{k}]  &  =\delta_{k[j}\mathbf{\bar{P}%
}_{i]}, & \left[  \mathbf{M,J}_{ij}\mathbf{,J}_{kl}\mathbf{,J}_{mn}\right]
&  =(3!)^{2}\alpha\delta_{\lbrack ij,kl,mn]}\mathbf{\bar{M},} \label{EBF1}%
\end{align}
where we define%
\begin{equation}
\delta_{\lbrack ij,kl,mn]}=(\delta_{ik}\delta_{jn}\delta_{lm}-\delta
_{jk}\delta_{in}\delta_{lm}-\delta_{il}\delta_{jn}\delta_{km}+\delta
_{jl}\delta_{in}\delta_{km}-\delta_{ik}\delta_{jm}\delta_{ln}+\delta
_{jk}\delta_{im}\delta_{ln}+\delta_{il}\delta_{jm}\delta_{kn}-\delta
_{jl}\delta_{im}\delta_{kn}).
\end{equation}
In this case, the higher-degree form is a three-form and consequently, all the
products in (\ref{EBF1}) are antisymmetric. Note that the first seven
relations describe the EB Lie algebra, while the products of the form $\left[
\mathbf{t}_{A}\mathbf{,\bar{t}}_{B}\right]  $ lie in the extended subspace but
inheriting the same structure constants. Furthermore, the information about
the FDA cocycle is contained in the last three multilinear products.

\subsection{Infinitesimal gauge transformations}

The use of $L_{\infty}$ algebras allows to write the components of FDA1 gauge
fields as linear combinations of vectors, such as it is usually done in the
study of Lie gauge theories. Let us consider the one-form $\mathbf{A}$ and a
three-form gauge field $\mathbf{B}$, evaluated in the standard and extended
sectors of the corresponding $L_{\infty}$ algebra, to whose components we
denote as
\begin{align}
\mathbf{A}  &  =\tau\mathbf{H}+m\mathbf{M}+e^{i}\mathbf{P}_{i}+\omega
^{i}\mathbf{G}_{i}+\frac{1}{2}\omega^{ij}\mathbf{J}_{ij}+\frac{1}{2}%
s^{ij}\mathbf{S}_{ij},\label{ebA}\\
\mathbf{B}  &  =b\mathbf{\bar{H}}+c\mathbf{\bar{M}}+b^{i}\mathbf{\bar{P}}%
_{i}+c^{i}\mathbf{\bar{G}}_{i}+\frac{1}{2}b^{ij}\mathbf{\bar{J}}_{ij}+\frac
{1}{2}c^{ij}\mathbf{\bar{S}}_{ij}, \label{ebB}%
\end{align}
where $\tau$, $m$, $e^{i}$, $\omega^{i}$, $\omega^{ij}$ and $s^{ij}$ are
one-forms, while $b$, $c$, $b^{i}$, $c^{i}$, $b^{ij}$ and $c^{ij}$ are
three-forms. Let us also consider the gauging of EB-FDA by introducing the
following curvatures%
\begin{align}
\mathbf{R}\left(  \mathbf{A}\right)   &  =R\left(  \tau\right)  \mathbf{H+}%
R\left(  m\right)  \mathbf{M}+R\left(  e^{i}\right)  \mathbf{P}_{i}+\frac
{1}{2}R\left(  \omega^{ij}\right)  \mathbf{J}_{ij}+R\left(  \omega^{i}\right)
\mathbf{G}_{i}+\frac{1}{2}R\left(  s^{ij}\right)  \mathbf{S}_{ij},\label{RA}\\
\mathbf{R}\left(  \mathbf{B}\right)   &  =H\left(  b\right)  \mathbf{\bar{H}%
+}H\left(  c\right)  \mathbf{\bar{M}+}H\left(  c^{i}\right)  \mathbf{\bar{G}%
}_{i}+H\left(  b^{i}\right)  \mathbf{\bar{P}}_{i}+\frac{1}{2}H\left(
b^{ij}\right)  \mathbf{\bar{J}}_{ij}+\frac{1}{2}H\left(  c^{ij}\right)
\mathbf{\bar{S}}_{ij}, \label{RB}%
\end{align}
where $\mathbf{R}\left(  \mathbf{A}\right)  $ is a two-form evaluated in the
Lie subalgebra EB, while $\mathbf{R}\left(  \mathbf{B}\right)  $ is a
four-form evaluated in the remaining subspace. The components of
$\mathbf{R}\left(  \mathbf{A}\right)  $ and $\mathbf{R}\left(  \mathbf{B}%
\right)  $ are explicitly given by eqs. (\ref{mceb1}) and (\ref{mceb2})
respectively (notice that they are no longer vanishing). Moreover, we
introduce the following gauge parameters evaluated in both algebraic sectors:
\begin{align}
\boldsymbol{\theta}  &  =\theta^{A}\mathbf{t}_{A}=\theta\mathbf{H}+\theta
^{i}\mathbf{P}_{i}+\frac{1}{2}\theta^{ij}\mathbf{J}_{ij}+\sigma\mathbf{M}%
+\sigma^{i}\mathbf{G}_{i}+\frac{1}{2}\sigma^{ij}\mathbf{S}_{ij},
\label{param1}\\
\mathbf{\Theta}  &  =\Theta^{A}\mathbf{\bar{t}}_{A}=\Theta\mathbf{\bar{H}%
}+\Theta^{i}\mathbf{\bar{P}}_{i}+\frac{1}{2}\Theta^{ij}\mathbf{\bar{J}}%
_{ij}+\Sigma\mathbf{\bar{M}}+\Sigma^{i}\mathbf{\bar{G}}_{i}+\frac{1}{2}%
\Sigma^{ij}\mathbf{\bar{S}}_{ij}. \label{param2}%
\end{align}
Here, $\theta$, $\theta^{i}$, $\theta^{ij}$, $\sigma$, $\sigma^{i}$ and
$\sigma^{ij}$ are zero-forms evaluated in the Lie subalgebra EB, while
$\Theta$, $\Theta^{i}$, $\Theta^{ij}$, $\Sigma$, $\Sigma^{i}$ and $\Sigma
^{ij}$ are two-forms evaluated in the extended sector. The infinitesimal gauge
variations of the gauge fields are given by
\cite{Castellani:1995gz,Castellani:2005vt,Castellani:2013mka,Castellani:2006jg}
\begin{align}
\delta A^{A}  &  =\mathrm{d}\theta^{A}+C_{BC}^{A}A^{B}\theta^{C},\\
\delta B^{I}  &  =\mathrm{d}\Theta^{I}+C_{AJ}^{I}A^{A}\Theta^{J}-C_{AJ}%
^{I}\theta^{A}B^{J}-\frac{1}{p!}C_{A_{1}\cdots A_{p+1}}^{I}\theta^{A_{1}%
}A^{A_{2}}\cdots A^{A_{p+1}}.
\end{align}
These variations can be equivalently written in terms of the dual $L_{\infty}$
basis as%
\begin{align}
\delta\mathbf{A}  &  =\mathrm{d}\boldsymbol{\theta}+\left[  \mathbf{A}%
,\boldsymbol{\theta}\right]  ,\label{deltaA2}\\
\delta\mathbf{B}  &  =\mathrm{d}\mathbf{\Theta}+\left[  \mathbf{A}%
,\mathbf{\Theta}\right]  -\left[  \boldsymbol{\theta}\mathbf{,B}\right]
-\frac{1}{\left(  p!\right)  ^{2}}\left[  \boldsymbol{\theta},\mathbf{A}%
,\ldots,\mathbf{A}\right]  . \label{deltaB2}%
\end{align}
In FDA gauge theory, these infinitesimal gauge variations induce definitions
of covariant derivatives for zero-forms and two-forms, such that the
variations can be written as $\delta\mathbf{A=\nabla}\boldsymbol{\theta}$ and
$\delta\mathbf{B=\nabla\Theta}$. Then, by consider the gauge fields from eqs.
(\ref{ebA}) and (\ref{ebB}) and the gauge parameters from eqs (\ref{param1})
and (\ref{param2}), and plugging in the $L_{\infty}$ products (\ref{EBF1}), we
find the following infinitesimal gauge variations.

Variations of the one-forms:%
\begin{align}
\delta\tau &  \equiv\nabla\theta=\mathrm{d}\theta,\nonumber\\
\delta e^{i}  &  \equiv\nabla\theta^{i}=\mathrm{D}_{\omega}\theta^{i}%
+\omega^{i}\theta-\tau\sigma^{i}-\theta_{\text{ \ }j}^{i}e^{j},\nonumber\\
\delta m  &  \equiv\nabla\sigma=\mathrm{d}\sigma+\omega^{i}\theta_{i}%
-e^{i}\sigma_{i},\nonumber\\
\delta\omega^{i}  &  \equiv\nabla\sigma^{i}=\mathrm{D}_{\omega}\sigma
^{i}-\theta_{\text{ \ }j}^{i}\omega^{j},\nonumber\\
\delta\omega^{ij}  &  \equiv\nabla\theta^{ij}=\mathrm{D}_{\omega}\theta
^{ij},\nonumber\\
\delta s^{ij}  &  \equiv\nabla\sigma^{ij}=\mathrm{D}_{\omega}\sigma
^{ij}+\omega^{i}\sigma^{j}-\omega^{j}\sigma^{i}-\theta_{\text{ \ }k}^{i}%
s^{kj}+\theta_{\text{ \ }k}^{j}s^{ki}. \label{cov16}%
\end{align}

Variations of the three-forms:%
\begin{align}
\delta b  &  \equiv\nabla\Theta=\mathrm{d}\Theta-\theta\omega_{ij}%
\omega_{\text{ \ }k}^{i}\omega^{kj}+3\alpha\tau\theta_{ij}\omega_{\ \ k}%
^{i}\omega^{kj},\nonumber\\
\delta b^{i}  &  \equiv\nabla\Theta^{i}=\mathrm{D}_{\omega}\Theta^{i}%
-\Theta_{\text{ \ }j}^{i}e^{j}-\tau\Sigma^{i}+\omega^{i}\Theta+c^{i}%
\theta+b_{\text{ \ }j}^{i}\theta^{j}-\theta_{\text{ \ }j}^{i}b^{j}-\sigma
^{i}b\nonumber\\
&  -\alpha\theta^{h}\omega_{ij}\omega_{\text{ \ }k}^{i}\omega^{kj}+3\alpha
e^{h}\theta_{ij}\omega_{\ \ k}^{i}\omega^{kj},\nonumber\\
\delta b^{ij}  &  \equiv\nabla\Theta^{ij}=\mathrm{D}_{\omega}\Theta
^{ij}-\theta_{\text{ \ }k}^{i}b^{kj}+\theta_{\text{ \ }k}^{j}b^{ki}%
,\nonumber\\
\delta c  &  \equiv\nabla\Sigma=\mathrm{d}\Sigma-e^{i}\Sigma_{i}+\omega
^{i}\Theta^{j}\delta_{ij}+c^{i}\theta_{i}-b^{i}\sigma_{i}-\sigma\omega
_{ij}\omega_{\text{ \ }k}^{i}\omega^{kj}+3\alpha m\theta_{ij}\omega
_{\ \ k}^{i}\omega^{kj},\nonumber\\
\delta c^{i}  &  \equiv\nabla\Sigma^{i}=\mathrm{d}\Sigma^{i}-\Theta_{\text{
\ }j}^{i}\omega^{j}+\omega_{\text{ \ }j}^{i}\Sigma^{j}-\frac{1}{2}%
\theta_{\text{ \ }j}^{i}c^{j}+b_{\text{ \ }j}^{i}\sigma^{j},\nonumber\\
\delta c^{ij}  &  \equiv\nabla\Sigma^{ij}=\mathrm{D}_{\omega}\Sigma
^{ij}+\omega^{i}\Sigma^{j}-\omega^{j}\Sigma^{i}-\sigma^{i}c^{j}+\sigma
^{j}c^{i}+s_{\text{ \ }k}^{i}\Theta^{kj}-s_{\text{ \ }k}^{j}\Theta
^{ki}\nonumber\\
&  -\theta_{\text{ \ }k}^{i}c^{kj}+\theta_{\text{ \ }k}^{j}c^{ki}+b_{\text{
\ }k}^{i}\sigma^{kj}-b_{\text{ \ }k}^{j}\sigma^{ki}. \label{cov712}%
\end{align}

\section{Non-relativistic gravity}

In this section we aim to construct gauge invariant theories for the 4D EB
algebra, and the 5D EB-FDA. In general, it is possible to write down
Chern-Simons actions for specific cases of FDA1 in even and odd
dimensionalities. However, the presence of the cocycle in the algebra changes
the properties of the invariant tensor of its Lie subalgebra. Hence, when
constructing a Chern-Simons action, it becomes necessary to find new invariant
tensors. This is the main obstacle when writing down a gauge invariant action
principle for a given FDA1. Therefore, we consider an alternative approach,
namely, the formalism of non-linear realizations, which allows to extend the
gauge invariance of a gauge theory from a stability subalgebra $\mathfrak{h}$
to a larger algebra $\mathfrak{g}$
\cite{Stelle:1979aj,Ivanov:1981wn,Ivanov:1981wm,Grignani:1991nj}. A well-known
example of this procedure consists in the writing of four-dimensional general
relativity as a gauge theory of the Poincar\'{e} group. In that case, the full
algebra is the 4D Poincar\'{e} algebra $\mathfrak{g}=\mathfrak{iso}\left(
3,1\right)  =\mathrm{span}\left\{  \mathbf{P}_{a}\mathbf{,J}_{ab}\right\}  $,
while the stability subalgebra is chosen as the Lorentz algebra $\mathfrak{h}%
=\mathfrak{so}\left(  3,1\right)  =\mathrm{span}\left\{  \mathbf{J}%
_{ab}\right\}  $. The theory is formulated as follows: one considers the
one-form gauge connection evaluated in the full algebra
\begin{equation}
\mathbf{A}=e^{a}\mathbf{P}_{a}+\frac{1}{2}\omega^{ab}\mathbf{J}_{ab},
\end{equation}
and a zero-form multiplet evaluated in the coset space resulting from the full
algebra and the stability subalgebra $\boldsymbol{\phi}=\phi^{a}\mathbf{P}%
_{a}$, i.e., in the translational subspace spanned by $\left\{  \mathbf{P}%
_{a}\right\}  $. A secondary gauge connection is then introduced, as follows%
\begin{equation}
\mathbf{\tilde{A}}=e^{-\boldsymbol{\phi}}\mathrm{d}e^{\boldsymbol{\phi}%
}+e^{-\boldsymbol{\phi}}\mathbf{A}e^{\boldsymbol{\phi}}.
\end{equation}
The new connection is obtained by means of a gauge transformation of
$\mathbf{A}$ where the multilet $\boldsymbol{\phi}$ plays the role of gauge
parameter. However, this is not regarded as a symmetry transformation. The
connection $\mathbf{\tilde{A}}$, sometimes referred as non-linear gauge field
because of its non-linear dependence on the multiplet, plays the role of the
fundamental field of the theory. It is possible to prove that any quantity
constructed with $\mathbf{A}$ that is invariant under the transformations of
the stability subalgebra, will become invariant under the full algebra if
$\mathbf{A}$ is replaced by $\mathbf{\tilde{A}}$, as long as the components of
the gauge multiplet transform in an appropriate manner. Thus, by directly
calculating the functional form of the non-linear connection $\mathbf{\tilde
{A}}=\tilde{e}^{a}\mathbf{P}_{a}+\frac{1}{2}\tilde{\omega}^{ab}\mathbf{J}%
_{ab}$, one finds
\begin{equation}
\tilde{e}^{a}=e^{a}+\mathrm{D}_{\omega}\phi,\text{ \ \ \ \ }\tilde{\omega
}^{ab}=\omega^{ab}.
\end{equation}
As mentioned, the components of the non-linear connection are considered as
fundamental fields, which means that $\tilde{e}^{a}$ is considered the
vierbein of the corresponding gravity theory while $\omega^{ab}$ plays the
role of spin connection. Then, the four-dimensional Einstein-Hilbert action in
first order formalism can be written as%
\begin{equation}
I\left[  \mathbf{\tilde{A}}\right]  =\kappa\int\epsilon_{abcd}R\left(
\omega^{ab}\right)  \tilde{e}^{c}\tilde{e}^{d}, \label{EH}%
\end{equation}
where $\kappa$ is a dimensional constant. It is easy to verify that, under an
infinitesimal gauge transformation with parameter $\boldsymbol{\theta}%
=\theta^{a}\mathbf{P}_{a}+\frac{1}{2}\theta^{ab}\mathbf{J}_{ab}$, the
components of $\mathbf{A}$ transform according to $\delta e^{a}=\mathrm{D}%
_{\omega}\theta^{a}-\theta_{\text{ \ }b}^{a}e^{b}$ and $\delta\omega
^{ab}=\mathrm{D}_{\omega}\theta^{ab}$. These variations, together with the
additional transformation law of the zero-form multiplet
\begin{equation}
\delta\phi^{a}=-\theta^{a}+\theta_{\text{ \ }b}^{a}\phi^{b}, \label{mult}%
\end{equation}
lead to a translation invariant vierbein $\delta\tilde{e}^{a}=-\theta_{\text{
\ }b}^{a}\tilde{e}^{b}$. As a consequence, the Einstein-Hilbert action,
formulated in terms of $\mathbf{\tilde{A}}$ instead of $\mathbf{A}$, extends
its gauge invariance from the Lorentz algebra to the Poincar\'{e} algebra
\cite{Salgado:2001bn,Salgado:2003rf}. It is important to note that fixing the
multiplet transformation law according to eq. (\ref{mult}) is an essential
factor in order to obtain a symmetry enhancement in the action (\ref{EH}).
However, the multiplet does not explicitly appear in the action. It plays the
role of a Goldstone field and has no dynamics.

\subsection{Non-linear realization}

Let us now consider the same approach for the construction of gauge invariant
theories for the EB algebra and for EB-FDA. In order to apply the formalism of
non-linear realizations, we consider the following decomposition of the gauge
algebra. We choose EB-FDA as the full gauge algebra, while the stability
algebra is chosen as the spatial rotation algebra $\mathfrak{so}\left(
D-1\right)  \,$\ spanned by $\left\{  \mathbf{J}_{ij}\right\}  $. Contrary to
the relativistic case discussed above, we exclude the Lorentz boosts from the
stability algebra and consider only the subspace of spatial rotations. In the
relativistic case, the Lagrangian density is chosen as the Einstein-Hilbert
Lagrangian, which is a Lorentz invariant. On the other hand, the EB Lie
algebra does not contain the Lorentz algebra as a subalgebra, since when
deriving the EB algebra from the Poincar\'{e} algebra, it is necessary to
decompose the Lorentz index into spatial and temporal components, and perform
a non-relativistic expansion (or an In\"{o}n\"{u}-Wigner contraction in the
case of the Galilei algebra). These procedures preserve the spatial rotation
subalgebra but alter the structure of the Lorentz boost sector. This
modification is a natural consequence of the non-Lorentzian nature of
non-relativistic limits. The speed of light is considered as infinite in such
regimes, which removes the physical interpretation of the Lorentz boost
composing the remaining subspace. Therefore, the choice of $\mathfrak{so}%
\left(  D-1\right)  $ is consistent with the idea of constructing a
non-relativistic theory. Moreover, from a mathematical point of view, it turns
out to be convenient to work with spatial rotation invariants when
constructing the Lagrangian densities as the invariant tensors associated with
these algebras are well-known. As an alternative, it would also be possible to
consider the full EB Lie algebra as stability algebra. This approach would
reduce the dimensionality of the coset space resulting from EB-FDA and the EB
algebra. However, at the same time, it would increase the difficulty in
finding an invariant action principle. On the other hand, it would be possible
to consider a lower-dimensional algebra as stability algebra, namely, a
subalgebra of the spatial rotation algebra. This would not only increase the
number of components of the multiplet, but would also break the spatial
rotation covariance of the theory. For the purposes of this work, we will
consider the spatial rotation algebra as the stability subalgebra, which is
the residual symmetry of the Lorentz algebra that remains as a closed
structure after performing the non-relativistic expansion. Consequently, we
introduce a multiplet taking values on the coset space resulting from EB-FDA
and $\mathfrak{so}\left(  D-1\right)  $. Since we are dealing with a FDA1, the
parameter is composed by:%
\begin{align}
\boldsymbol{\phi} &  =\phi^{A}\mathbf{t}_{A}=\phi\mathbf{H}+\phi^{i}%
\mathbf{P}_{i}+\lambda\mathbf{M}+\lambda^{i}\mathbf{G}_{i}+\frac{1}{2}%
\lambda^{ij}\mathbf{S}_{ij},\\
\mathbf{\Phi} &  =\Phi^{A}\mathbf{\bar{t}}_{A}=\Phi\mathbf{\bar{H}}+\Phi
^{i}\mathbf{\bar{P}}_{i}+\frac{1}{2}\Phi^{ij}\mathbf{\bar{J}}_{ij}%
+\Lambda\mathbf{\bar{M}}+\Lambda^{i}\mathbf{\bar{G}}_{i}+\frac{1}{2}%
\Lambda^{ij}\mathbf{\bar{S}}_{ij}.
\end{align}
Here, $\phi$, $\phi^{i}$, $\lambda$, $\lambda^{i}$ and $\lambda^{ij}$ are
zero-forms, and $\Phi$, $\Phi^{i}$, $\Phi^{ij}$, $\Lambda$, $\Lambda^{i}$ and
$\Lambda^{ij}$ are two-forms. Note that $\boldsymbol{\phi}$ \ and
$\mathbf{\Phi}$ carry components along all the vectors of the dual
$L_{\infty}$ algebra, except those spanning the stability subalgebra. Then, we
introduce the non-linear gauge fields $\mathbf{\tilde{A}}$ and $\mathbf{\tilde
{B}}$ as gauge transformed fields of $\mathbf{A}$ and $\mathbf{B}$
respectively. The gauge transformation is carried out with $\boldsymbol{\phi}$
and $\mathbf{\Phi}$ playing the role of gauge parameters.%
\begin{equation}
\mathbf{\tilde{A}}=\mathbf{A}+\delta\mathbf{A},\text{ \ \ \ \ }\mathbf{\tilde
{B}}=\mathbf{B}+\delta\mathbf{B}.
\end{equation}
Although the multiplets are the parameters of a gauge transformation, this is
not considered a symmetry transformation of the theory. The action principle
will be written as a functional of $\mathbf{\tilde{A}}$ and $\mathbf{\tilde
{B}}$, whose infinitesimal transformation law will be fixed by the independent
transformations of $\left(  \mathbf{A},\mathbf{B}\right)  $ and $\left(
\boldsymbol{\phi},\mathbf{\Phi}\right)  $. We denote the components of the
non-linear fields as%
\begin{align}
\mathbf{\tilde{A}} &  =\tilde{\tau}\mathbf{H}+\tilde{m}\mathbf{M}+\tilde
{e}^{i}\mathbf{P}_{i}+\tilde{\omega}^{i}\mathbf{G}_{i}+\frac{1}{2}%
\tilde{\omega}^{ij}\mathbf{J}_{ij}+\frac{1}{2}\tilde{s}^{ij}\mathbf{S}_{ij},\\
\mathbf{\tilde{B}} &  =\tilde{b}\mathbf{H}+\tilde{c}\mathbf{M}+\tilde{b}%
^{i}\mathbf{P}_{i}+\tilde{c}^{i}\mathbf{G}_{i}+\frac{1}{2}\tilde{b}%
^{ij}\mathbf{J}_{ij}+\frac{1}{2}\tilde{c}^{ij}\mathbf{S}_{ij}.
\end{align}
The components of the one-form $\mathbf{\tilde{A}}$ are explicitly given by%
\begin{align}
\tilde{\tau} &  =\tau+\nabla\phi,\nonumber\\
\tilde{m} &  =m+\nabla\lambda-\frac{1}{2}\lambda_{i}\nabla\phi^{i}+\frac{1}%
{2}\phi_{i}\nabla\lambda^{i}+\frac{1}{6}\lambda^{2}\nabla\phi-\frac{1}{6}%
\phi\lambda_{i}\nabla\lambda^{i},\nonumber\\
\tilde{e}^{i} &  =e^{i}+\nabla\phi^{i}-\frac{1}{2}\lambda^{i}\nabla\phi
+\frac{1}{2}\phi\nabla\lambda^{i},\nonumber\\
\tilde{\omega}^{i} &  =\omega^{i}+\nabla\lambda^{i},\nonumber\\
\tilde{\omega}^{ij} &  =\omega^{ij},\nonumber\\
\tilde{s}^{ij} &  =s^{ij}+\nabla\lambda^{ij},\label{nleb1}%
\end{align}
with $\lambda^{2}=\lambda^{i}\lambda_{i}$. These are the components of the
non-linear gauge connection of EB, which is a natural consequence of the
presence of the EB Lie algebra as subalgebra of EB-FDA. Moreover, the
components of the non-linear three-form gauge field are given by
\begin{align}
\tilde{b} &  =b+\nabla\Phi,\nonumber\\
\tilde{b}^{i} &  =b^{i}+\nabla\Phi^{i}-\frac{1}{2}\lambda^{i}\nabla\Phi
-\frac{1}{2}\Lambda^{i}\nabla\phi+\frac{1}{2}\Phi\nabla\lambda^{i}+\frac{1}%
{2}\phi\nabla\Lambda^{i}-\frac{1}{2}\Phi_{\text{ \ }j}^{i}\nabla\phi^{i}%
+\frac{1}{2}\nabla\Phi^{ij}\phi_{j}\nonumber\\
&  -\frac{1}{3}\phi\Phi_{\text{ \ }j}^{i}\nabla\lambda^{j}+\frac{1}{6}%
\phi\nabla\Phi^{ij}\lambda_{j}+\frac{1}{6}\nabla\phi\Phi_{\text{ \ }j}%
^{i}\lambda^{j},\nonumber\\
\tilde{b}^{ij} &  =b^{ij}+\nabla\Phi^{ij},\nonumber\\
\tilde{c} &  =c+\nabla\Lambda-\frac{1}{2}\lambda_{i}\nabla\Phi^{i}-\frac{1}%
{2}\Lambda_{i}\nabla\phi^{i}+\frac{1}{2}\phi_{i}\nabla\Lambda^{i}+\frac{1}%
{2}\Phi_{i}\nabla\lambda^{i}-\frac{1}{3}\nabla\Phi^{ij}\lambda_{i}\phi
_{j}+\frac{1}{3}\lambda_{i}\Lambda^{i}\nabla\phi+\frac{1}{6}\lambda^{2}%
\nabla\Phi\nonumber\\
&  -\frac{1}{6}\Phi\lambda_{i}\nabla\lambda^{i}-\frac{1}{6}\phi\lambda
_{i}\nabla\Lambda^{i}+\frac{1}{6}\Phi_{ij}\lambda^{i}\nabla\phi^{j}-\frac
{1}{6}\Phi_{ij}\phi^{i}\nabla\lambda^{j}-\frac{1}{6}\phi\Lambda_{i}%
\nabla\lambda^{i}+\frac{1}{12}\phi\Phi_{ij}\lambda^{i}\nabla\lambda
^{j},\nonumber\\
\tilde{c}^{i} &  =c^{i}+\nabla\Lambda^{i}-\frac{1}{2}\Phi_{\text{ \ }j}%
^{i}\nabla\lambda^{j}-\frac{1}{2}\nabla\Phi^{ij}\Theta\lambda_{j},\nonumber\\
\tilde{c}^{ij} &  =c^{ij}+\nabla\Lambda^{ij}-\left[  \frac{1}{2}\lambda
^{i}\nabla\Lambda^{j}+\frac{1}{2}\Lambda^{i}\nabla\lambda^{j}+\frac{1}{2}%
\Phi_{\text{ \ }k}^{i}\nabla\lambda^{kj}-\frac{1}{2}\nabla\Phi^{ik}\lambda
_{k}^{\text{ \ }j}-\frac{1}{3}\lambda^{i}\Phi_{\text{ \ }k}^{j}\nabla
\lambda^{k}\right.  \nonumber\\
&  \left.  -\frac{1}{6}\Phi_{\text{ \ }k}^{i}\lambda^{k}\nabla\lambda
^{j}+\frac{1}{6}\lambda^{i}\nabla\Phi^{jk}\lambda_{k}-\left(  i\leftrightarrow
j\right)  \right]  .\label{nleb2}%
\end{align}
Note that we have written the non-linear gauge fields by using the definition
of covariant derivative introduced in eqs. (\ref{cov16}) and (\ref{cov712}).
The components of the non-linear gauge curvature two-form $\mathbf{R}\left(
\mathbf{\tilde{A}}\right)  $ can be obtained from the general expression in
eqs. (\ref{mceb1}) i.e.,%
\begin{align}
R(\tilde{\tau}) &  =\mathrm{d}\tilde{\tau},\nonumber\\
R(\tilde{m}) &  =\mathrm{d}\tilde{m}+\tilde{\omega}^{i}\tilde{e}%
_{i},\nonumber\\
R(\tilde{e}^{i}) &  =\mathrm{D}_{\omega}\tilde{e}^{i}+\tilde{\omega}^{i}%
\tilde{\tau},\nonumber\\
R(\tilde{\omega}^{ij}) &  =R(\omega^{ij}),\nonumber\\
R(\tilde{\omega}^{i}) &  =\mathrm{D}_{\omega}\tilde{\omega}^{i},\nonumber\\
R(\tilde{s}^{ij}) &  =\mathrm{D}_{\omega}\tilde{s}^{ij}+\tilde{\omega}%
^{i}\tilde{\omega}^{j}.
\end{align}
Since the algebra has been gauged, these curvatures are not vanishing.
Similarly, the components of the non-linear curvature four-form $\mathbf{R}%
\left(  \mathbf{\tilde{B}}\right)  $ are obtaining by replacing the components
of $\mathbf{\tilde{A}}$ and $\mathbf{\tilde{B}}$ into eqs. (\ref{mceb2}%
).\ They are explicitly given by:%
\begin{align}
R\left(  \tilde{b}\right)   &  \equiv\mathrm{d}\tilde{b}+\alpha\tilde{\tau
}\tilde{\omega}^{ij}\tilde{\omega}_{ik}\tilde{\omega}_{\text{ \ }j}%
^{k},\nonumber\\
R\left(  \tilde{c}\right)   &  \equiv\mathrm{d}\tilde{c}+\tilde{\omega}%
^{i}\tilde{b}_{i}+\tilde{b}^{i}\tilde{e}_{i}+\alpha\tilde{m}\tilde{\omega
}^{ij}\tilde{\omega}_{ik}\tilde{\omega}_{\text{ \ }j}^{k},\nonumber\\
R\left(  \tilde{b}^{i}\right)   &  \equiv\mathrm{D}_{\omega}\tilde{b}%
^{i}+\tilde{b}\tilde{\omega}^{i}+\tilde{c}^{i}\tilde{\tau}+\tilde{b}_{\text{
\ }j}^{i}\tilde{e}^{j}+\alpha\tilde{e}^{i}\tilde{\omega}^{jk}\tilde{\omega
}_{jl}\tilde{\omega}_{\text{ \ }k}^{l},\nonumber\\
R\left(  \tilde{c}^{i}\right)   &  \equiv\mathrm{D}_{\omega}\tilde{c}%
^{i}+\tilde{b}_{\text{ \ }j}^{i}\tilde{\omega}^{j},\nonumber\\
R\left(  \tilde{b}^{ij}\right)   &  \equiv\mathrm{D}_{\omega}\tilde{b}%
^{ij},\nonumber\\
R\left(  \tilde{c}^{ij}\right)   &  \equiv\mathrm{D}_{\omega}\tilde{c}%
^{ij}+\tilde{s}_{\text{ \ }k}^{i}\tilde{b}^{kj}-\tilde{s}_{\text{ \ }k}%
^{j}\tilde{b}^{ki}+\tilde{\omega}^{i}\tilde{c}^{j}-\tilde{\omega}^{j}\tilde
{c}^{i}.\label{curvB}%
\end{align}
At this point, we have obtained the infinitesimal gauge variations of
components of $\mathbf{A}$, and the functional form of the non-linear gauge
fields as well. Since the fundamental fields of the theory are $\mathbf{\tilde
{A}}$ and $\mathbf{\tilde{B}}$, the transformation law of the multiplet must
also be specified. As before, this transformation law is a key ingredient in
order to enlarge the symmetry the action principle that we will propose. The
transformation law of the multiplet can be found in Appendix A.

\subsection{Action principles}

Eqs. (\ref{nleb1}) show the functional form of the non-linear gauge fields of
the $D$-dimensional EB Lie algebra. It is therefore possible to write down a
gauge invariant action principle for EB by proposing a functional of
$\mathbf{A}$ that presents off-shell gauge invariance under spatial rotations.
Then, the components of $\mathbf{A}$ must be replaced by those of
$\mathbf{\tilde{A}}$. This will extend the gauge invariance from the spatial
rotation subalgebra to the full EB algebra. For this purpose, let us first
recall that an action principle for Newtonian gravity\ in arbitrary dimensions
has been introduced in second order formalism in ref. \cite{Hansen:2019pkl}.
Moreover, in refs. \cite{Bergshoeff:2016lwr,Ekiz:2022wbi}, the 4D theory has
been presented in first order formulation as a gauge invariant theory of the
Newtonian algebra \cite{Ozdemir:2019tby}. This 4D action has been obtained by
means of a power series expansion of bimetric gravity. In our case, we also
set $D=4$ and consider the following functional%
\begin{equation}
I^{\text{EB}}\left[  \mathbf{\tilde{A}}\right]  =\kappa\int\epsilon
_{ijk}\left(  R\left(  \tilde{\omega}^{ij}\right)  \tilde{\tau}\tilde{e}%
^{k}+R\left(  \tilde{\omega}^{i}\right)  \tilde{e}^{j}\tilde{e}^{k}+R\left(
\tilde{s}^{ij}\right)  \tilde{\tau}\tilde{e}^{k}+R\left(  \tilde{\omega}%
^{ij}\right)  \tilde{m}\tilde{e}^{k}\right)  . \label{action1}%
\end{equation}
The first term of inside the integral in eq. (\ref{action1}) corresponds to
Galilean gravity, which, although it does not lead to the Poisson equation by
itself, it represents the most simple consistent non-relativistic gravity
model \cite{Cariglia:2018hyr,Hansen:2020wqw,Guerrieri:2020vhp} (see ref.
\cite{Bergshoeff:2017btm} for second-order formulation). Moreover, eq.
(\ref{action1}) contains the terms that are present in the aforementioned 4D
Newtonian action. The variation of the full action principle\ leads to the
following equations of motion%
\begin{align}
\delta\tilde{\tau}  &  :\text{ \ \ }\epsilon_{ijk}\left(  R(\tilde{\omega
}^{ij})\tilde{e}^{k}+R(\tilde{s}^{ij})\tilde{e}^{k}\right)  =0,\label{d4em1}\\
\delta\tilde{e}^{i}  &  :\text{ \ \ }\epsilon_{ijk}\left(  R(\tilde{\omega
}^{jk})\tilde{\tau}+2\mathrm{D}_{\omega}\tilde{\omega}^{j}\tilde{e}%
^{k}+R(\tilde{s}^{jk})\tilde{\tau}+R(\tilde{\omega}^{jl})\tilde{m}\right)
=0,\label{d4em2}\\
\delta\tilde{\omega}^{i}  &  :\text{ \ \ }\epsilon_{ijk}(\tilde{\omega}%
^{j}\tilde{\tau}\tilde{e}^{k}+\mathrm{D}_{\omega}\tilde{e}^{j}\tilde{e}%
^{k})=\epsilon_{ijk}R(\tilde{e}^{j})\tilde{e}^{k},\label{d4em3}\\
\delta\tilde{\omega}^{ij}  &  :\text{ \ \ }\epsilon_{ijk}\left(  \tilde{e}%
^{k}\mathrm{d}\left(  \tilde{\tau}+\tilde{m}\right)  -\mathrm{D}_{\omega
}\tilde{e}^{k}\left(  \tilde{\tau}+\tilde{m}\right)  \right)  -\epsilon
_{imk}\left(  2\tilde{s}_{\text{ \ }j}^{m}\tilde{\tau}-\tilde{\omega}%
_{j}\tilde{e}^{m}\right)  \tilde{e}^{k}=0,\label{d4em4}\\
\delta\tilde{m}  &  :\text{ \ \ }\epsilon_{ijk}R(\tilde{\omega}^{ij})\tilde
{e}^{k}=0,\label{d4em5}\\
\delta\tilde{s}^{ij}  &  :\text{ \ \ }\epsilon_{ijk}(\mathrm{d}\tilde{\tau
}\tilde{e}^{k}-\tilde{\tau}\mathrm{D}_{\omega}\tilde{e}^{k})=0. \label{d4em6}%
\end{align}
Note that eq. (\ref{d4em3}) leads to $R(\tilde{e}^{i})=0$ as field equation.
Moreover, eq. (\ref{d4em5}) shows that the gauge curvature $R(\tilde{\omega
}^{ij})$ is on-shell vanishing, which implies that the hypersurfaces are flat.
It is then possible to introduce the following ansatz for the spatial
components of the spin connection in terms of the gravitational potential
$\tilde{\omega}^{i}=-\partial^{i}\phi\left(  x\right)  \mathrm{d}x^{0}$ (see
ref. \cite{Concha:2022jdc}). Then, by replacing eq. (\ref{d4em1}) into
(\ref{d4em2}) one finds that the gravitational potential satisfies the
standard Poisson equation $\nabla^{2}\phi(x)=0$. Thus, the writing of
(\ref{action1}) as a functional of the non-linear gauge fields from eqs.
(\ref{nleb1}), allows a formulation of Newtonian gravity as a genuine
off-shell gauge invariant theory of the EB algebra. This is similar to what
occurs when gauging the Newtonian algebra in the ordinary approach. However,
it must be noticed that there is a fundamental difference between
(\ref{action1}) and the mentioned 4D Newtonian action, since the latter
carries an extra gauge field which is necessary for a gauge invariant
formulation. Thus, in order to preserve gauge invariance, the formulation of
such Newtonian action makes use of a higher-dimensional algebra and
consequently it carries a larger number of gauge fields. In contrast, the
action (\ref{action1}) does not require enlarging either the EB gauge symmetry
or the number of gauge fields, but to extend the standard EB field content by
means of the introduction of\ the coset space multiplet. Although both
formulations are fundamentally different from a mathematical point of view,
they share the feature of extending the EB field content to preserve gauge
invariance and leading to the Poisson equation as resulting dynamics. The 3D
version of the action principle (\ref{action1}) has been studied within the
framework of Chern-Simons theories within the first and second order
formalisms
\cite{Papageorgiou:2009zc,Bergshoeff:2016lwr,Ozdemir:2019tby,Hartong:2016yrf}
and recently derived as a contraction of the Einstein-Hilbert action that
preserves the invariance of the non-relativistic Lagrangians in a certain
scaling limit \cite{Bergshoeff:2016lwr,Ekiz:2022wbi}.

Let us now consider the use of the full non-linear realization of EB-FDA,
given by eqs. (\ref{nleb1}) and (\ref{nleb2}). As before, we identify the
stability subalgebra as the spatial rotation algebra. In this case, we must
propose a Lagrangian density fulfilling the requirement of being a spatial
rotation scalar and a functional of $\mathbf{\tilde{A}}$ and $\mathbf{\tilde
{B}}$. This requires the inclusion of kinetic terms for the three-forms. Since
the extended gauge curvatures (\ref{curvB}) are given by four-forms, the
dimensionality in which a consistent action principle can be found is
restricted as $D\geq5$. Such a requirement prevents the construction of a
three- or four-dimensional theory based on EB-FDA using this procedure. Such a
limitation could be circumvented by studying the existence of
Chevalley-Eilenberg cohomology classes of the EB Lie algebra (and other
non-relativistic symmetries). For example, the finding of a three-cocycle as a
replacement for the four-cocycle (\ref{O2}) would lead to a theory coupling
one-forms with two-forms, which in turn would imply constructing a Lagrangian
density consisting of three-forms curvatures instead of the four-forms
presented here. It is important to recall that the existence of cohomology
classes not only depends on the dimensionality of the cocycles but also on the
chosen representation (for examples of this kind in the context of
non-relativistic algebras, see \cite{Bonanos:2008kr}). In this case, we have
considered the construction of a four-cocycle in the adjoint representation of
the EB Lie algebra. However, the study and classification of these
cohomologies in other representations, dimensionalities, and for other
non-relativistic symmetries is an open problem.

In this case, we set $D=5$ and propose the following functional:%
\begin{align}
I^{\text{EB-FDA}}\left[  \mathbf{\tilde{A},\tilde{B}}\right]   &  =\kappa%
%TCIMACRO{\dint }%
%BeginExpansion
{\displaystyle\int}
%EndExpansion
\epsilon_{ijkl}\left(  R(\omega^{ij})\tilde{e}^{k}\tilde{e}^{l}\tilde{\tau
}+\frac{2}{3}R(\tilde{\omega}^{i})\tilde{e}^{j}\tilde{e}^{k}\tilde{e}%
^{l}+R(\tilde{s}^{ij})\tilde{e}^{k}\tilde{e}^{l}\tilde{\tau}+R(\omega
^{ij})\tilde{e}^{k}\tilde{e}^{l}\tilde{m}\right) \nonumber\\
&  +\tau R(\tilde{c})-\tilde{m}\left(  R(\tilde{b})+R(\tilde{c})\right)
+\left(  \tilde{\omega}_{i}+\tilde{e}_{i}\right)  R(\tilde{b}^{i}).
\label{action2}%
\end{align}
Here, $\kappa$ is a dimensional constant. The Lagrangian density inside the
integral contains the one-forms of the EB Lie algebra, which present a
geometric interpretation, in addition to the contributions depending on the
four-form gauge curvatures. In order to obtain a non-relativistic gravity
interpretation for this action, we split the Lagrangian in sectors depending
on the one-forms of the EB Lie algebra and those terms depending on the
components of curvature four-form. Thus, we denote%
\begin{align}
L_{\mathbf{\tilde{A}}}  &  =\epsilon_{ijkl}\left(  R(\omega^{ij})\tilde{e}%
^{k}\tilde{e}^{l}\tilde{\tau}+\frac{2}{3}R(\tilde{\omega}^{i})\tilde{e}%
^{j}\tilde{e}^{k}\tilde{e}^{l}+R(\tilde{s}^{ij})\tilde{e}^{k}\tilde{e}%
^{l}\tilde{\tau}+R(\omega^{ij})\tilde{e}^{k}\tilde{e}^{l}\tilde{m}\right)
,\label{LG}\\
L_{\mathbf{\tilde{B}}}  &  =\tilde{\tau}R(\tilde{c})-\tilde{m}\left(
R(\tilde{b})+R(\tilde{c})\right)  +\left(  \tilde{\omega}_{i}+\tilde{e}%
_{i}\right)  R(\tilde{b}^{i}). \label{LM}%
\end{align}
Thus, we introduce $L_{\mathbf{\tilde{A}}}$ as a pure gravity Lagrangian. This
is the 5D version of the one presented in eq. (\ref{action1}). Consequently,
its variation leads to the standard 5D Poisson equation. It must be noticed
that, in order to present gauge covariance, $L_{\mathbf{\tilde{B}}}$ is
written in terms of the gauge curvatures. However, the cocycle in eqs.
(\ref{curvB}) introduces terms of geometric nature in $L_{\mathbf{\tilde{B}}}%
$. These terms do not depend on the three-forms but only on the one-forms of
the EB algebra. Consequently, they could be included in $L_{\mathbf{\tilde{A}%
}}$. However, for the purposes of this work, we maintain these terms in the
second contribution. The equations of motion that emerge from the variations
of the Lagrangian with respect to the one-forms can be written as follows:%
\begin{align}
\frac{\delta L_{\mathbf{\tilde{A}}}}{\delta\tilde{\tau}}  &  =\epsilon
_{ijkl}\left(  R(\tilde{\omega}^{ij})\tilde{e}^{k}\tilde{e}^{l}+R(\tilde
{s}^{ij})\tilde{e}^{k}\tilde{e}^{l}\right)  =-\ast\mathcal{T}_{0}%
,\label{em1}\\
\frac{\delta L_{\mathbf{\tilde{A}}}}{\delta\tilde{e}^{i}}  &  =2\epsilon
_{ijkl}\left(  R(\tilde{\omega}^{jk})\tilde{e}^{l}\tilde{\tau}-\mathrm{D}%
_{\omega}\tilde{\omega}^{j}\tilde{e}^{k}\tilde{e}^{l}+R(\tilde{s}^{jk}%
)\tilde{e}^{l}\tilde{\tau}+R(\tilde{\omega}^{jl})\tilde{e}^{l}\tilde
{m}\right)  =-\ast\mathcal{T}_{i},\label{em2}\\
\frac{\delta L_{\mathbf{\tilde{A}}}}{\delta\tilde{\omega}^{i}}  &
=2\epsilon_{ijkl}(\tilde{\omega}^{j}\tilde{\tau}\tilde{e}^{k}\tilde{e}%
^{l}+\mathrm{D}_{\omega}\tilde{e}^{j}\tilde{e}^{k}\tilde{e}^{l})=-\ast
\mathcal{S}_{i},\label{em3}\\
\frac{\delta L_{\mathbf{\tilde{A}}}}{\delta\tilde{\omega}^{ij}}  &
=\epsilon_{ijkl}\left(  \tilde{e}^{k}\tilde{e}^{l}\mathrm{d}\left(
\tilde{\tau}+\tilde{m}\right)  +2\mathrm{D}_{\omega}\tilde{e}^{k}\tilde{e}%
^{l}\left(  \tilde{\tau}+\tilde{m}\right)  \right)  -2\epsilon_{imkl}\left(
\tilde{s}_{\text{ \ }j}^{m}\tilde{\tau}-\frac{1}{3}\tilde{\omega}_{j}\tilde
{e}^{m}\right)  \tilde{e}^{k}\tilde{e}^{l}\nonumber\\
&  =-\ast\mathcal{S}_{ij},\label{em4}\\
\frac{\delta L_{\mathbf{\tilde{A}}}}{\delta\tilde{m}}  &  =\epsilon
_{ijkl}R(\tilde{\omega}^{ij})\tilde{e}^{k}\tilde{e}^{l}=\mathrm{d}\tilde
{b}+\mathrm{d}\tilde{c}+\tilde{\omega}^{i}\tilde{b}_{i}+\tilde{b}^{i}\tilde
{e}_{i}+2\alpha\tilde{\tau}\tilde{\omega}^{ij}\tilde{\omega}_{ik}\tilde
{\omega}_{\text{ \ }j}^{k},\label{em5}\\
\frac{\delta L_{\mathbf{\tilde{A}}}}{\delta\tilde{s}^{ij}}  &  =\epsilon
_{ijkl}(\mathrm{d}\tilde{\tau}\tilde{e}^{k}\tilde{e}^{l}-2\tilde{\tau
}\mathrm{D}_{\omega}\tilde{e}^{k}\tilde{e}^{l}). \label{em6}%
\end{align}
\newline Here, we introduce $\mathcal{T}_{0}$ and $\mathcal{T}_{i}$ as the
time- and space-components of an energy-momentum one-form. Similarly, we
introduce $\mathcal{S}_{ij}$ and $\mathcal{S}_{i}$ as the components of a
relativistic spin one-form. These components are defined in terms of the
functional variations of $L_{\mathbf{\tilde{B}}}$ with respect to the same
fields, as follows:%
\begin{align}
\frac{\delta L_{\mathbf{\tilde{B}}}}{\delta\tilde{\tau}}  &  =\ast
\mathcal{T}_{0}=\mathrm{d}\tilde{c}+\tilde{\omega}^{i}\tilde{b}_{i}+\tilde
{b}^{i}\tilde{e}_{i}+\tilde{\omega}_{i}\tilde{c}^{i}+\tilde{e}_{i}\tilde
{c}^{i}+2\alpha\tilde{m}\tilde{\omega}^{ij}\tilde{\omega}_{ik}\tilde{\omega
}_{\text{ \ }j}^{k},\label{matter1}\\
\frac{\delta L_{\mathbf{\tilde{B}}}}{\delta\tilde{e}^{i}}  &  =\ast
\mathcal{T}_{i}=\tilde{\tau}\tilde{b}_{i}-\tilde{m}\tilde{b}_{i}+\tilde
{b}_{ij}\tilde{\omega}^{j}+\mathrm{D}_{\tilde{\omega}}\tilde{b}_{i}+\tilde
{b}\tilde{\omega}_{i}+\tilde{c}_{i}\tilde{\tau}+2\tilde{b}_{ij}\tilde{e}%
^{j}-\alpha\tilde{\omega}_{i}\tilde{\omega}^{jk}\tilde{\omega}_{jl}%
\tilde{\omega}_{\text{ \ }k}^{l},\label{matter2}\\
\frac{\delta L_{\mathbf{\tilde{B}}}}{\delta\tilde{\omega}^{i}}  &
=\ast\mathcal{S}_{i}=\mathrm{D}_{\omega}\tilde{b}_{i}-\tilde{\tau}\tilde
{b}_{i}+\tilde{m}\tilde{b}_{i}+\tilde{e}_{i}\tilde{b}+\tilde{c}_{i}\tilde
{\tau}+\tilde{b}_{ij}\tilde{e}^{j}+\alpha\tilde{e}_{i}\tilde{\omega}%
^{jk}\tilde{\omega}_{jl}\tilde{\omega}_{\text{ \ }k}^{l},\label{matter3}\\
\frac{\delta L_{\mathbf{\tilde{B}}}}{\delta\tilde{\omega}^{ij}}  &
=\ast\mathcal{S}_{ij}=-\tilde{e}_{i}\tilde{b}_{j}-\tilde{\omega}_{i}\tilde
{b}_{j}+6\alpha\tilde{\tau}\tilde{m}\tilde{\omega}_{ik}\tilde{\omega}_{\text{
\ }j}^{k}+\alpha\tilde{\omega}_{l}\tilde{e}^{l}\tilde{\omega}_{ik}%
\tilde{\omega}_{\text{ \ }j}^{k}. \label{matter4}%
\end{align}
Eqs. (\ref{matter1})-(\ref{matter4}) show that the three-form gauge fields are
source of non-relativistic curvature and torsion. However, by setting these
three-forms as zero, the dynamics of standard Newtonian gravity is not
immediately recovered. Interestingly, the presence of the cocycle in the
definition of the gauge curvatures modifies the vacuum of the theory in a way
in which the geometric fields show a different behavior even in the absence of
three-forms. In fact, eqs. (\ref{em1})-(\ref{em4}) show that, by turning off
the three-forms and considering $\alpha=1$, the theory still presents
non-vanishing curvature and torsion. However, eqs. (\ref{matter1}%
)-(\ref{matter4}) show that this is a consequence of the mentioned purely
geometric terms in $L_{\mathbf{\tilde{B}}}$. This establishes a connection to
the five-dimensional version of the Einstein-Cartan-Sciama-Kibble (ECSK)
theory \cite{adamowicz1975equivalence,adamowicz1975principle}, in which the
Lagrangian density $L_{\mathbf{\tilde{B}}}$ plays the role of an effective
matter Lagrangian. Thus, the vacuum of the theory behaves as the standard 5D
non-relativistic theory\ (whose Lagrangian density is given by (\ref{LG})) in
the presence of effective matter and spin densities. To examine this scenario,
let us fix an orthonormal set of coordinates, and consider the following
ansatz for the spatial and temporal non-zero components of the spin connection
in orthonormal basis \cite{Castagnino,Castagnino2}:%
\begin{align}
\tilde{\omega}^{i}  &  =-\partial^{i}\phi\mathrm{d}x^{0},\\
\tilde{\omega}^{ij}  &  =\frac{1}{2}\left(  \delta_{l}^{i}\partial^{j}%
\phi-\delta_{l}^{j}\partial^{i}\phi+\text{\texttt{s}}_{\text{ }l}^{j}%
v^{i}+\text{\texttt{s}}^{ij}v_{l}+\text{\texttt{s}}_{\text{ }l}^{i}%
v^{j}\right)  \mathrm{d}x^{l}.
\end{align}
Here $\phi\left(  x\right)  $ is the gravitational potential. Moreover we
introduce an effective spin density \texttt{s}$^{ij}\left(  x\right)  $ with
spatial velocity $v^{i}$. Note that we are not adding a new spin Lagrangian
into the theory. On the contrary, we are considering an ansatz that includes
the functions \texttt{s}$^{ij}\left(  x\right)  $ in order to describe the
intrinsic non-vanishing torsion of the theory. These functions are related
with the effective spin density one-form defined in eqs. (\ref{em4}) and
(\ref{matter4}) according to
\begin{equation}
\mathcal{S}_{l}^{ij}=\text{\texttt{s}}^{ij}v_{l},
\end{equation}
and verify \texttt{s}$^{ij}=-$\texttt{s}$^{ji}$ and \texttt{s}$_{\text{ \ }%
j}^{i}v^{j}=0$. Thus, the ansatz relates the components of $\tilde{\omega}%
^{i}$ and $\tilde{\omega}^{ij}$ in the chosen orthonormal system with the
components of the effective spin density \cite{RevModPhys.48.393}. Next, we
set the $\tilde{m}$ field to zero, replace the ansatz in the equations of
motion, plug in eq. (\ref{em1}) into eq. (\ref{em2}), and compute its trace in
tensorial language. This procedure leads to the following modified Poisson
equation%
\begin{equation}
\nabla^{2}\phi(x)+\frac{1}{6}\epsilon_{ijkl}v^{2}\partial^{i}\phi
\text{\texttt{s}}_{\text{ \ }m}^{j}\text{\texttt{s}}_{\text{ \ }n}%
^{m}\text{\texttt{s}}^{nk}v^{l}=0, \label{Poisson}%
\end{equation}
with $v^{2}=v^{i}v_{i}$. The presence of the second term in the l.h.s. of eq.
(\ref{Poisson}) is a direct consequence of the cocycle in the gauge algebra.
It is noteworthy that the EB algebra was extended to a FDA by means of its own
topological properties and, the non-triviality of the cocycle $\Omega_{2}$
implies a richer gauge structure in which the three-forms cannot be decomposed
as combinations of one-forms. It might be expected that, by turning off the
three-forms, we would recover vanishing spatial torsion and the standard
Poisson equation as dynamics. This would be the case if no cocycle were
considered in the construction of the FDA. Moreover, if the cocycle were
trivial, it could be reabsorbed into the three-forms by a redefinition of the
gauge fields. Consequently, the non-triviality of the cocycle $\Omega_{2}$ is
the origin of a modified dynamics with on-shell non-vanishing curvature and
torsion. Thus, standard Newtonian dynamics, with vanishing spatial curvature
and torsion, and flat hypersurfaces in the corresponding foliation is
recovered by setting the three-forms as vanishing, in addition to $\alpha=0$.

It would be interesting to explore the second order formalism of the 5D action
proposed in eq. (\ref{action2}). Since its variation leads to a modified
Newtonian dynamics, it would be possible to make a connection with the 5D case
of the Newtonian action principle proposed in ref. \cite{Hansen:2019pkl}.
However, it must be noticed that such formulation is in general more
challenging due to the presence of the higher-degree forms and the cocycle in
the equations of motion. Moreover, it would also be possible to explore the
potential connection with the 5D Chern-Simons actions that emerge from the
gauging of non-relativistic symmetries, especially in the case of the 5D EB
algebra, where the field content would match the one presented here, at least
for the sector spanned by the one-forms.

\section{Concluding remarks}

In this work, we have introduced a novel FDA as an extension of the EB Lie
algebra. This FDA includes three-forms in the adjoint representation of the EB
algebra, and consequently, its gauging leads to an extended field content.
This approach highlights the role of Chevalley-Eilenberg cohomology classes in
the construction of a gravity theory. In this context, we have made use of the
formalism of non-linear realizations to formulate standard 4D Newtonian
gravity as a gauge theory of the EB algebra. Furthermore, we have proposed a
5D gauge theory for non-relativistic gravity in the presence of three-forms.
Regarding the 5D theory, the presence of the cocycle (representative of the
cohomology class) in the construction of the algebra generates terms of
geometric nature in the extended sector of the Lagrangian, which is
constructed with the components of the higher-degree curvature. Then, we have
derived the equations of motion and approached the problem in the framework of
ECSK gravity, by turning off the higher-degree fields, and proposing an ansatz
for the geometric ones in terms of the gravitational potential and an
effective spin density. The replacement of this ansatz in the equations of
motion leads to a modification of the Poisson equation. Although this approach
is useful for studying the effective behavior of the non-relativistic
gravitational field, it is important to clarify that it describes the vacuum
state of the theory. In other words, the theory presents intrinsic on-shell
non-zero curvature and torsion. The behavior of such torsion is studied by
adding the effective spin density in the ansatz. In this way, the theory can
be studied as a standard 5D non-relativistic gravity, in the presence of a
non-zero spin density.

It is important to point out that, although FDAs first appeared in the
formulation of higher-dimensional supergravity, their mathematical consistency
does not require supersymmetry and can be defined in purely bosonic contexts,
as it is the case with EB-FDA. It is noteworthy that in ref.
\cite{Bonanos:2008kr}, the Chevalley-Eilenberg cohomology classes of the
bosonic Galilei and Poincar\'{e} algebras have been studied. These results
were carried out in a different representation than the one studied in this
work, which makes them inequivalent to the cohomology class associated with
EB-FDA. As future work, it would be interesting to formulate the
non-relativistic FDAs and gravity theories that emerge from these cohomology
classes. Additionally, it would be interesting to study compactifications of
the theories presented in this work. This would allow, for example, to find 4D
gravity theories that couple one-forms with three-forms. It would then be
possible to study the gauge symmetries presented by these theories in order to
find a potential connection with new FDAs presenting non-trivial cocycles in
lower dimensions and different representations. Moreover, it is noteworthy
that there are several isomorphisms between relativistic, non-relativistic and
ultra relativistic algebras. For example, the three-dimensional Poincar\'{e}
and EB algebras are isomorphic to the so-called AdS-Carroll and extended
AdS-static algebras \cite{Bacry:1968zf,Bacry:1970ye,Concha:2023bly}. These are
changes of basis wherein different physical interpretations are assigned to
the gauge fields and symmetry generators. Since the existence of non-trivial
cocycles do not depend on the chosen basis, it would be possible to construct
FDAs for Carrollian algebras and study the effect of including cocycles in
their resulting dynamics. Moreover, it would be also interesting to explore
the existence of supersymmetric extensions of the cocycle presented in eq.
(\ref{O2}) and the mentioned ones from ref. \cite{Bonanos:2008kr}. This would
allow to formulate non-relativistic supergravity models in higher dimensions
and their possible connection with relativistic supergravity theories that
naturally include higher-degree forms in their field content. In this way,
bosonic FDA gauge theories (such as the one presented in this work) can be
proposed as the bosonic sector of supergravity theories with $p$-forms. A
similar approach was considered in ref. \cite{Camarero:2017yka}, where a
bosonic Chern-Simons model with $p$-forms was proposed as the bosonic sector
of standard eleven dimensional supergravity.%

\section*{Acknowledgements}

A.M. is supported by the ANID Postdoctoral Fellowship No. 74240083.

%TCIMACRO{\TeXButton{\appendix}{\appendix}}%
%BeginExpansion
\appendix
%EndExpansion

\section{Transformation law of the multiplet}

Let us consider an infinitesimal gauge transformation of EB-FDA. The gauge
parameter is composed by the zero-form and two-form from eqs. (\ref{param1})
and (\ref{param2}). As it was mentioned, within the formalism of non-linear
realizations, the infinitesimal transformation law of the multiplet is set in
a way that allows the extension of the gauge symmetry. For details on the
formal definition of these gauge variations, and their applications to
specific cases of FDA1, see refs.
\cite{Stelle:1979aj,Ivanov:1981wn,Ivanov:1981wm,Grignani:1991nj,Salgado:2021sjo,Salgado:2023owk}%
. In this case, we find that the components of the multiplet one-form
$\boldsymbol{\phi}$ transform according to:%
\begin{align}
\delta\phi &  =-\theta,\nonumber\\
\delta\phi^{i}  &  =-\theta^{i}-\theta_{\text{ \ }j}^{i}\phi^{j}-\frac{1}%
{2}\lambda^{i}\theta+\frac{1}{2}\phi\sigma^{i},\nonumber\\
\delta\lambda &  =-\sigma-\theta_{\text{ \ }j}^{i}\lambda^{j}+\frac{1}{2}%
\phi^{i}\sigma_{i}-\frac{1}{2}\lambda^{i}\theta_{i}-\frac{1}{12}\lambda
^{2}\theta+\frac{1}{12}\phi\lambda_{i}\sigma^{i},\nonumber\\
\delta\lambda^{i}  &  =-\sigma^{i},\nonumber\\
\delta\lambda^{ij}  &  =-\sigma^{ij}-\theta_{\text{ \ }k}^{i}\lambda
^{kj}+\theta_{\text{ \ }k}^{j}\lambda^{ki}-\frac{1}{2}\lambda^{i}\sigma
^{j}+\frac{1}{2}\lambda^{j}\sigma^{i}. \label{mult2}%
\end{align}
Eqs. (\ref{mult2}) are the non-relativistic counterpart of the transformation
law of the Poincar\'{e} multiplet in eq. (\ref{mult}). The derivation of these
transformation laws completes the gauging of the EB Lie algebra in the
formalism of non-linear realizations.

On the other hand, the components of the multiplet two-form $\mathbf{\Phi}$
transform according to:%
\begin{align}
\delta\Phi &  =-\Theta+3\alpha\phi\theta_{ij}\omega_{\text{ \ }k}^{i}%
\omega^{kj},\nonumber\\
\delta\Phi^{i}  &  =-\Theta^{i}-\theta_{\text{ \ }j}^{i}\Phi^{j}+3\alpha
\phi^{i}\theta_{jk}\omega_{\text{ \ }l}^{j}\omega^{lk}+3\alpha\lambda
\theta_{ij}\omega_{\text{ \ }k}^{i}\omega^{kj}+\frac{1}{2}\lambda^{i}%
\Theta+\frac{1}{2}\Lambda^{i}\theta-\frac{1}{2}\Phi\sigma^{i}-\frac{1}{2}%
\phi\Sigma^{i}\nonumber\\
&  +\frac{1}{2}\Phi_{\text{ \ }j}^{i}\theta^{j}-\frac{1}{2}\Theta_{\text{
\ }j}^{i}\phi^{j}+\frac{1}{6}\phi\Phi_{\text{ \ }j}^{i}\sigma^{j}-\frac{1}%
{12}\phi\Theta_{\text{ \ }j}^{i}\lambda^{j}-\frac{1}{12}\theta\Phi_{\text{
\ }j}^{i}\lambda^{j},\nonumber\\
\delta\Phi^{ij}  &  =-\Theta^{ij}-\theta_{\text{ \ }k}^{i}\Phi^{kj}%
+\theta_{\text{ \ }k}^{j}\Phi^{ki},\nonumber\\
\delta\Lambda &  =-\Sigma+\frac{1}{2}\lambda^{i}\Theta_{i}-\frac{1}{2}\Phi
^{i}\sigma_{i}-\frac{1}{2}\phi^{i}\Sigma_{i}+\frac{1}{2}\Lambda^{i}\theta
_{i}-\frac{1}{6}\theta\Lambda_{i}\lambda^{i}+\frac{1}{6}\Theta_{ij}\lambda
^{i}\phi^{j}+\frac{1}{12}\phi\Lambda_{i}\sigma^{i}\nonumber\\
&  -\frac{1}{12}\lambda^{2}\Theta+\frac{1}{12}\Phi\lambda_{i}\sigma^{i}%
+\frac{1}{12}\phi\lambda_{i}\Sigma^{i}-\frac{1}{12}\Phi_{ij}\lambda^{i}%
\theta^{j}+\frac{1}{12}\Phi_{ij}\phi^{i}\sigma^{j},\nonumber\\
\delta\Lambda^{i}  &  =-\Sigma^{i}-\theta_{\text{ \ }j}^{i}\Lambda^{j}%
+\frac{1}{2}\Phi_{\text{ \ }j}^{i}\sigma^{j}-\frac{1}{2}\Theta_{\text{ \ }%
j}^{i}\lambda^{j},\nonumber\\
\delta\Lambda^{ij}  &  =-\Sigma^{ij}-\theta_{\text{ \ }k}^{i}\Lambda
^{kj}+\theta_{\text{ \ }k}^{j}\Lambda^{ki}+\frac{1}{2}\left(  \lambda
^{i}\Sigma^{j}+\Lambda^{i}\sigma^{j}+\Phi_{\text{ \ }k}^{i}\sigma^{kj}%
-\Theta_{\text{ \ }k}^{i}\lambda^{kj}-\frac{1}{3}\lambda^{i}\Phi_{\text{ \ }%
k}^{j}\sigma^{k}\right. \nonumber\\
&  \left.  -\frac{1}{6}\Phi_{\text{ \ }k}^{i}\lambda^{k}\sigma^{j}+\frac{1}%
{6}\lambda^{i}\Theta_{\text{ \ }k}^{j}\lambda^{k}-\left(  i\leftrightarrow
j\right)  \right)  . \label{mult2B}%
\end{align}
The setting of the transformation laws (\ref{mult2B}), in addition to
(\ref{mult2}), completes the gauging of EB-FDA in the formalism of non-linear realizations.
\bibliographystyle{utphys.bst}
 
\bibliography{Template-FDANR}

\end{document}